\documentclass[12pt,preprint]{aastex}

\newcommand{\skipthis}[1]{}

\newcommand{\kms}{\hbox{km\,s$^{-1}$}}
\newcommand{\Msun}{\mbox{${\rm M}_{\sun}$}}
\newcommand{\Lsun}{\mbox{$L_{\sun}$}}
\newcommand{\cm}[1]{\mbox{cm$^{#1}$}}   
\newcommand{\cube}{\mbox{cm$^{-3}$}}  
\newcommand{\cc}{\mbox{cm$^{-2}$}}  

\newcommand{\nnh}{N$_2$H$^{+}$}
\newcommand{\nhhh}{NH$_3$}
\newcommand{\nnd}{N$_2$D$^{+}$}

\newcommand{\tex}{\mbox{$T_{\rm ex}$}}
\newcommand{\h}{\mbox{${\rm ^h}$}}
\newcommand{\m}{\mbox{${\rm ^m}$}}

\shortauthors{Bourke et al.}
\shorttitle{Oph A-N6 SMA}

\begin{document}

\title{Initial conditions for star formation in clusters: physical and
kinematical structure of the starless core OphA-N6}

\author{
Tyler L.\ Bourke\altaffilmark{1}, 
Philip C.\ Myers\altaffilmark{1}, 
Paola Caselli\altaffilmark{2}, 
James Di Francesco\altaffilmark{3},
Arnaud Belloche\altaffilmark{4},
Ren\'e Plume\altaffilmark{5},
David J. Wilner\altaffilmark{1} 
}

\altaffiltext{1}{Harvard-Smithsonian Center for Astrophysics, 60 Garden
Street, Cambridge, MA 02138; email tbourke@cfa.harvard.edu}
\altaffiltext{2}{School of Physics \& Astronomy,
E.C. Stoner Building, The University of Leeds, Leeds, LS2 9JT, UK}
\altaffiltext{3}{National Research Council Canada, Herzberg Institute of
Astrophysics, Victoria, BC, Canada}
\altaffiltext{5}{Max-Planck-Institut f\"ur Radioastronomie,
Auf dem H\"ugel 69, D-53121 Bonn, Germany}
\altaffiltext{4}{Department of Physics and Astronomy, University of
Calgary, Calgary, AB, Canada}

\begin{abstract}
We present high spatial ($<$300 AU) and spectral (0.07 \kms) resolution
Submillimeter Array observations of the dense starless cluster core Oph A
N6, in the 1 mm dust continuum and the 3-2 line of \nnh\ and \nnd.  The
dust continuum observations reveal a compact source not seen in single-dish
observations, of size $\sim$1000 AU and mass 0.005-0.01 \Msun.  The
combined line and single-dish observations reveal a core of size
3000 $\times$ 1400 AU elongated in a NW-SE direction, with almost no
variation in either line width or line center velocity across the map, and
very small non-thermal motions.  The deuterium fraction has a peak value of
$\sim$0.15 and is $>$ 0.05 over much of the core.  The \nnh\ column density
profile across the major axis of Oph A-N6 is well represented by 
an isothermal cylinder, with temperature 20 K, peak density $7.1 \times
10^6$ \cube, and \nnh\ abundance $2.7 \times 10^{-10}$.  The mass of Oph
A-N6 is estimated to be 0.29 \Msun, compared to a value of 0.18 \Msun\ from
the isothermal cylinder analysis, and 0.63 \Msun\ for the critical mass for
fragmentation of an isothermal cylinder.  Compared to isolated low-mass
cores, Oph A-N6 shows similar narrow line widths and small velocity
variation, with a deuterium fraction similar to ``evolved" dense cores.  It
is significantly smaller than isolated cores, with larger peak column and
volume density.  The available evidence suggests Oph A-N6 has formed
through the fragmentation of the Oph A filament and is the precursor to a
low-mass star.  The dust continuum emission suggests it may already
have begun to form a star.
\end{abstract}


\section{Introduction}
\label{sec-intro}

It is now well established that low-mass (i.e., solar-like) stars form from
the collapse of dense cores within molecular clouds (e.g., Larson 2003).
The initial conditions of isolated star formation are known from continuum
and line observations of many tens of dense cores in molecular cloud
complexes such as Taurus, with relatively sparse concentrations of young
stars (Di Francesco et al.\ 2007).  The best commonly observed molecular
line tracers of the central few thousand AU of centrally condensed cores on
the verge of star formation are \nhhh\ (ammonia), \nnh, and \nnd\ (Caselli
et al.\ 2002c; Tafalla et al.\ 2002; Crapsi et al.\ 2005, 2007).
Observations of these lines and the dust continuum have enabled the
properties of dense isolated starless cores thought to be near to or at the
point of star formation to be well determined in recent years (Di Francesco
et al.\ 2007; Bergin \& Tafalla 2007).  These properties include (i) a high
degree of deuterium fractionation, (i.e., large N(\nnd)/N(\nnh)) (ii) a
large central density ($\sim 10^6$ \cube), (iii) depletion of CO and other
C-bearing species, (iv) cold central regions ($<$10 K), and (v) line
asymmetries indicative of infall motions.

In this paper, we present the first detailed observational study of the
internal structure of a starless cluster core, to explore how star
formation in clusters compares to isolated star formation.  Most stars are
believed to form in close proximity to other stars within embedded clusters
(Lada \& Lada 2003; Allen et al.\ 2007; Bressert et al.\ 2010).  As a
result, a fundamental problem in astrophysics is whether star formation in
cluster environments is similar to better-understood isolated star
formation, or whether cluster star formation is more turbulent and dynamic.
An important question is to what degree do stars in clusters form in the
same way as in isolation, i.e., from cores whose properties strongly
influence the stellar properties (Shu et al.\ 2004; Larson 2005; Tan et
al.\ 2006; see reviews by Shu et al.\ 1987; Larson 2003), and to what
degree do they form in a different, more dynamic way, where external forces
and interactions matter more than initial conditions (e.g., Bonnell et al.\
2001; Bonnell \& Bate 2006).  

A number of observational studies made over recent years 
have provided information on the global properties of dense cores within
nearby cluster-forming regions (Ward-Thompson et al.\ 2007).  Dust
continuum observations tracing high column densities have been more
numerous, due to their faster mapping speeds compared to molecular line
observations.  As a result, large maps have been made of molecular clouds
containing embedded clusters in Ophiuchus (L1688; Motte et al.\ 1998;
Johnstone et al.\ 2000; Stanke et al.\ 2006; Young et al.\ 2006), Perseus
(NGC 1333, IC348-SW; Sandell \& Knee 2001; Hatchell et al.\ 2005; Enoch et
al.\ 2006; H.~Kirk et al.\ 2006), Corona Australis (Chini et al.\ 2003;
Nutter et al.\ 2005), Serpens (Davis et al.\ 1999; Enoch et al.\ 2007), and
the more distant and massive Orion cluster (Chini et al.\ 1997; Lis et al.\
1998; Johnstone \& Bally 1999; Li et al.\ 2007).  Some progress has been
made in large area mapping of molecular line dense gas tracers with
resolution approaching that of dust continuum observations of $\leq
30\arcsec$, in nearby low-mass regions (Williams \& Myers 2000; Andr{\'e}
et al.\ 2007; Walsh et al.\ 2007; Friesen et al.\ 2009, 2010a,b), and Orion
(Ikeda et al.\ 2007, 2009; Tatematsu et al.\ 2008).  Although some studies
have been able to ``resolve'' starless cores in clusters in molecular
spectral lines (Williams \& Myers 2000; Walsh et al 2007), in the sense
that the measured core size is larger than the beam, no study has examined
the internal structure of any cluster core in detail.  Observations of
cores in cluster-forming regions that resolve individual cores are thus
needed to understand how these regions can make stars so much more
efficiently than in isolation.  Progress toward this goal has been slower
than for isolated regions.  Most young clusters are more crowded so their
cores suffer more confusion; their cores are smaller and more distant so
they are harder to resolve, and they form stars more frequently so that
starless cores are less common (Jijina et al.\ 1999). 

\subsection{Ophiuchus A N6}

The Ophiuchus molecular cloud, at a distance of only 125 pc (de Geus et
al.\ 1989; Knude \& Hog 1998; Loinard et al.\ 2008; Lombardi et al.\ 2008),
is the nearest example of cluster formation embedded within dense gas and
dust, and is an ideal region in which to study the initial conditions for
cluster formation (see review by Wilking, Gagn\'e and Allen 2008; Motte et
al.\ 1998; Johnstone et al.\ 2000, 2004; Andr\'e et al. 2007; Enoch et al.\
2008; J{\o}rgensen et al.\ 2008; Simpson et al.\ 2008; Padgett et al.\
2008; Friesen et al. 2009, 2010a,b; van Kempen et al.\ 2009; Gutermuth et
al.\ 2009; Maruta et al.\ 2010).  Within Ophiuchus, the Oph A ridge is the
brightest of the clumps in both dust emission and \nnh\ 1-0, containing 8
dust cores identified throught 1.3 mm continuum emission (Motte et al.\
1998) and 6 local maxima of integrated \nnh\ 1-0 emission (Di Francesco et
al.\ 2004, hereafter DAM04; Andr\'e et al.\ 2007).  The \nnh\ cores in the
main part of the ridge, where the dust emission is strongest, show line
widths that are not significantly different from those observed toward
isolated starless cores (Di Francesco et al.\ 2004; Crapsi et al.\ 2005).
In this region, N6 is the best core for studies of internal structure as it
is isolated from the other Oph A cores (thereby suffering less from
confusion), is bright in molecular lines, and is larger and thus better
resolved than the other Oph A cores (Di Francesco et al.\ 2004; Pon et al.\
2009).  For these reasons, we have undertaken high angular resolution
submillimeter observations of N6 using the Submillimeter Array in the high
density gas tracers \nnh\ and \nnd\ 3-2, and combined those data with
single dish observations using the James Clerk Maxwell Telescope (JCMT) and
Institut de Radioastronomie Millim\'etrique (IRAM) 30-m telescope.

This paper is divided into the following sections.  In \S2 we present our
observations using the Submillimeter Array (SMA), JCMT, and IRAM 30-m, in
\S3 we present the line and continuum results, \S4 presents the analysis,
including the column densities, deuterium fraction, and structure of N6,
\S5 presents a discussion on the structure and evolution of N6, and
compares it to both isolated cores and cores within cluster-forming
regions, while a summary is presented in \S6.

\section{Observations}

\subsection{Submillimeter Array}

Observations in the \nnd\ 3-2 line toward Oph A-N6 were undertaken with the
SMA on 2006 July 22.  The array was in its compact configuration and zenith
opacities at 225 GHz were typically 0.1-0.13.  The SMA correlator was
configured with 2048 channels over 104 MHz for the \nnd\ 3-2 line at
231.321 GHz, providing a channel spacing of 0.066 \kms.  This high
resolution mode decreases the available bandwidth for continuum
observations, resulting in 1 GHz of continuum bandwidth in the upper
sideband and 1.15 GHz in the lower.  The effective continuum frequency is
226 GHz (1.3 mm).  The observations of Oph A N6 were interleaved with the
quasar J1626-298 for complex gain calibration.  Uranus and 3c454.3 were
used for bandpass calibration, and Uranus was also used for flux
calibration.  The data were calibrated and edited in the SMA's MIR software
package.  

Observations in the \nnh\ 3-2 line toward Oph A-N6 were undertaken with the
SMA on 2007 March 30.  The array was in its compact configuration and
zenith opacities at 225 GHz were typically $\sim0.06$.  The SMA correlator
was configured with 2048 channels over 104 MHz for the \nnh\ 3-2 line at
279.512 GHz, providing a channel spacing of 0.055 \kms.  This high
resolution mode decreases the available bandwidth for continuum
observations, resulting in 1.3 GHz of continuum bandwidth in both the upper
and lower sidebands.  The effective continuum frequency is 276
GHz (1.1 mm).  The observations of Oph A-N6 were interleaved with the
quasars J1626-298 and J1517-243 for complex gain calibration.  The quasar
3c279 was used for bandpass calibration, and Titan and 3c279 were used for
flux calibration.  The data were calibrated and edited in the SMA's MIR
software package.  

Further observations in the \nnh\ 3-2 line toward Oph A-N6 were undertaken
with the SMA on 2007 May 2, with the array in its subcompact configuration.
Zenith opacities at 225 GHz were typically 0.06-0.08.  The correlator
setup, resolution, available continuum bandwidth, and complex gain
calibrators were the same as for the 2007 March 30 observations.  Neptune
and 3c273 were used for bandpass calibration, and Neptune was also used for
flux calibration.  The data were calibrated and edited in the SMA's MIR
software package.  

Based upon independent observations of the gain and passband quasars at
similar frequencies ($\pm10$ GHz) and times (within one month), and upon
independent calibration of the observations presented here, as part of the
SMA's ongoing monitoring of quasar fluxes\footnote{see
http://sma1.sma.hawaii.edu/callist/callist.html}, we estimate the flux
calibration to be good to 20\% for all our observations.

\subsection{JCMT}

Observations in the \nnd\ 3-2 line toward Oph A-N6 were undertaken with the
James Clerk Maxwell Telescope (JCMT) between February and July 2005.  All
observations were made on a $5\times5$ grid with 10\arcsec\ spacing, with
angular resolution of 22\arcsec, and a spectral resolution of 0.1 \kms.
Details of the observations have been presented in Pon et al.\ (2009), and
the reader is referred to that paper for further information.

\subsection{IRAM 30-m}

Observations of Oph A-N6 with the IRAM 30-m telescope were undertaken in
May 2007 as part of a larger mapping program of the Oph A ridge.  Although
four SIS heterodyne receivers were used simultaneously in the 3, 2, and 1.2
mm atmospheric windows, here we only focus on the observations of \nnh\ 3-2
at 279.511 GHz that are used in this paper.  The autocorrelation
spectrometer VESPA was used as backend with a channel spacings of 40~kHz
and bandwidth of 80~MHz.  The system temperature ranged from 650 to 3620~K
and the pointing was checked every 1--2 hours on bright quasars and found
to be good to 2--3\arcsec\ (rms).  The telescope focus was optimized on
Saturn and Jupiter every 3--4 hours.  At the frequency of \nnh\ 3-2 the
telescope beam size (full-width at half-power) is 9\arcsec.  The
observations were performed in position-switching mode with the OFF
position offset by ($\Delta\alpha$,$\Delta\delta$) = (-900\arcsec,0\arcsec)
from the nominal map center of 16\h26\m26\fs46,
-24\degr24\arcmin30.8\arcsec, located at VLA 1623.  No emission was found
at the OFF position down to an rms noise level of 0.42~K in $T_a^*$ scale.
Mapping was done in on-the-fly scanning mode with a step of 4$''$,
providing fully sampled maps. We scanned alternately in right ascension and
declination to avoid striping artefacts. The data were reduced using the
CLASS software in its Fortran 90 version\footnote{see
http://www.iram.fr/IRAMFR/GILDAS}

\subsection{Combined Interferometric and Single-Dish Observations}

The procedure used to combine the interferometric and single dish data sets
is similar to that described by Zhang et al. (2000) and
Takakuwa et al.\ (2007), and is based on the methods described in Vogel et
al. (1984) and Wilner \& Welch (1994).  The MIRIAD software package
(Sault et al.\ 1995) was used for the combination and subsequent imaging.   

\subsubsection{\nnd\ 3-2: SMA + JCMT}

For the \nnd\ 3-2 observations, the data sets were resampled along the
velocity axis to a channel spacing of 0.07 \kms.  The JCMT data were
converted to Jy using a conversion factor of $S(Jy) = 27.4 \times T_A^*
(K)$, and deconvolved with a 22\arcsec\ FWHM Gaussian used to represent the
JCMT beam at 231 GHz.  Next, the JCMT data were convolved by a 55\arcsec\
Gaussian representing the SMA primary beam (full width at half power).
Side lobe effects are not well known and are thus ignored.  Then the JCMT
image cube was fourier transformed into a visibility data set, with a
sampling density in the (u,v) plane chosen to closely match that of the SMA
in their overlap region.  Finally, the JCMT and SMA visibility data sets
were fourier transformed together back into the image plane.  Because of
the extended nature of the emission, a correction for the SMA primary beam
attenuation away from the phase center was applied.  With a robust
weighting of 0 applied during the transform, the resultant image cube has a
resolution of $5\farcs1 \times 3\farcs4$ (synthesised beam full-width at
half power) with a 1$\sigma$ rms sensitivity of 0.43 Jy beam$^{-1}$
channel$^{-1}$.

\subsubsection{\nnh\ 3-2: SMA + 30-m}

The procedure for combining the SMA and 30-m \nnh\ 3-2 data sets is the
same as that used for the SMA and JCMT data sets.  The data sets were first
resampled along the velocity axis to a channel spacing of 0.07 \kms, for
direct comparison with the \nnd\ 3-2 data.  A conversion factor of $S(Jy) =
9.3 \times T_A^* (K)$ was used for the 30-m data, and the assumed Gaussian
beams sizes used were 8\farcs8 and 45\arcsec\ for the 30-m and SMA
respectively at 279.5 GHz.  The 30-m and SMA  visibility data sets were
fourier transformed together using a robust weighting of 0 with a Gaussian
taper of 2\arcsec.  Because of the extended nature of the emission, a
correction for the SMA primary beam attenuation away from the phase center
was applied.  The resultant image cube has a resolution of $5\farcs6 \times
3\farcs7$, similar to that of the \nnd\ 3-2 cube, with a 1$\sigma$ rms
sensitivity of 0.72 Jy beam$^{-1}$ channel$^{-1}$.  A comparison of the
central spectra of the combined SMA + 30-m dataset with the 30-m only
dataset, after smoothing the SMA + 30-m dataset to the angular resolution
of the 30-m dataset, shows that essentially all of the single-dish flux is
recovered, and the line-shapes are very similar.  

\subsection{Continuum imaging}

The MIRIAD software package was used to fourier transform and produce
images from the interferometric-only continuum data.  At 226 GHz (1.3 mm) a
robust weighting of 2 was used with a 3\arcsec\ taper, to improve the
sensitivity, resulting in a resolution of $5\farcs2 \times 4\farcs3$ with a
1$\sigma$ rms sensitivity of 2.6 mJy beam$^{-1}$.  At 276 GHz (1.1 mm) a
robust weighting of 0 was used with a 3\arcsec\ taper (note that with both
compact and subcompact array data, a robust weighting of 2 would
excessively down-weight the longer baselines), resulting in a resolution of
$4\farcs6 \times 3\farcs5$ with a 1$\sigma$ rms sensitivity of 3.6 mJy
beam$^{-1}$.

\section{Results}

\subsection{Molecular line maps}

\begin{figure}[!t]
\centering
\includegraphics[height=3.5in,angle=-90]{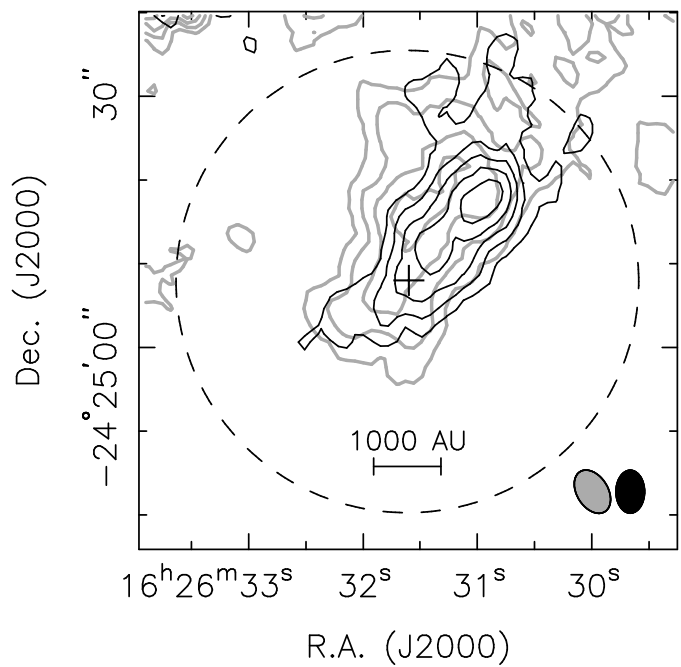}
\caption{Comparison of the integrated line maps of \nnd\ 3-2 (black
contours) and \nnh\ 3-2 (grey contours).  The black cross is the position
of the peak integrated \nnh\ 1-0 from DAM04. The synthesised beam size for
each observation is shown in the corresponding colours at lower right, and
the primary beam size for the \nnd\ observations is shown as the dashed
circle.  Contour levels are 10, 14, 18 ... times the $1\sigma$ sensitivity of
0.12 Jy/beam \kms\ for \nnd\ 3-2, and 12, 16, 20 ... times the $1\sigma$
sensitivity of 0.47 Jy/beam \kms\ for \nnh\ 3-2.
%
}
\label{map-compare-lines}
\end{figure}

Figure~\ref{map-n2dp32} shows the integrated line maps of \nnd\ 3-2 and
\nnh\ 3-2, compared to the distribution of integrated \nnh\ 1-0 emission
within the entire Oph A ridge (DAM04).  The integration is performed over
the hyperfine structure, corresponding to velocity ranges of 3.23--4.56
\kms\ (\nnd) and 0.22--6.18 \kms\ (\nnh).  The general agreement between
the \nnd\ and \nnh\ emission is good, but their peaks are offset by
9\arcsec\ (Figure~\ref{map-compare-lines}).  Similarly, the \nnh\ 1-0 map
has its peak offset from the \nnd\ 3-2 map, and shows good positional
coincidence with \nnh\ 3-2.  The offsets between \nnh\ and \nnd\ could be
due to optical depth effects (see \S\ref{sec-coldens}), or chemical
differentiation (Pon et al.\ 2009).  Figure~\ref{map-compare-lines} also
shows that the 3-2 maps have slightly different position angles.  The \nnd\
3-2 map is closely aligned with the \nnh\ 1-0 map, although their peaks do
not coincide.  

\begin{figure}[!t]
\centering
\includegraphics[height=6in,angle=-90]{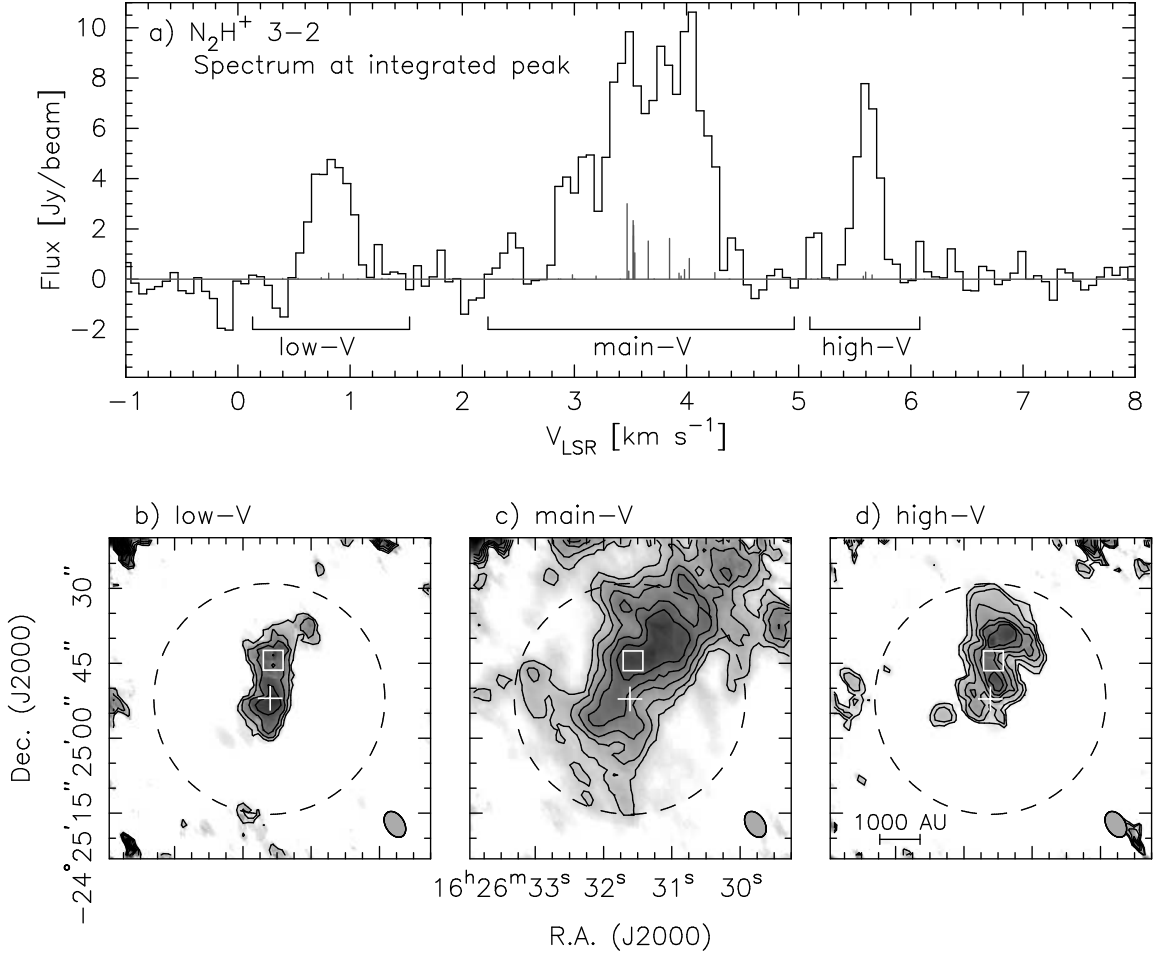}
\caption{Spectrum of \nnh\ 3-2, panel (a), and maps of integrated \nnh\ 3-2
emission over the 3 hyperfine groups, panels (b)-(d).  Data is combined
30-m + SMA (compact+subcompact).  In (a), the location of the hyperfine
components, and their relative weights, are indicated by the light grey
lines.  The limits of integration are indicated by ``low-V", ``main-V", and
``high-V", for panels (b)-(d).  The rms noise in the spectrum is 0.5
Jy/beam.  Contour levels are (b) 3, 4, 5, ... times
the $1\sigma$ sensitivity of 0.26 Jy/beam \kms, (c) 15, 18, 21, ... times
the $1\sigma$ sensitivity of 0.30 Jy/beam \kms, and (c) 3, 4, 5, ... times
the $1\sigma$ sensitivity of 0.21 Jy/beam.  The white cross is the position
of the peak integrated \nnh\ 1-0 emission (DAM04), while the white square
marks the position of the spectrum shown in (a).  The large dashed circle
indicates the size of the SMA primary beam.  The small grey ovals at lower
right in panels (b)-(d) indicate the synthesised beam size.
%
} 
\label{map-n2hp32-hfs} 
\end{figure}

The \nnh\ 3-2 spectrum is composed of three groups of hyperfine (hf)
features.  Integrated maps of these hyperfine groups are shown in
Figure~\ref{map-n2hp32-hfs}, labelled according to their velocity offsets
relative to the line frequency as ``low-V", ``main-V", and ``high-V".  The
low-V and high-V groups are sometimes referred to as the satellite
hyperfine groups.  Figure~\ref{map-n2hp32-hfs}(a) shows the spectrum at the
position of peak integrated emission in the main-V map, with the positions
and relative intensities of the hyperfine components in the optically thin
case indicated.  The main-V group shows
considerable saturation for the components in the range 3.5--4.0 \kms,
indicating large optical depths (confirmed through fitting of the hyperfine
structure, see \S\ref{sec-coldens}).  As a result, the integrated map of
the main-V group is larger than that of the other hyperfine groups and the
region of peak emission is more extended.  The peaks of the satellite
hyperfine groups, low-V and high-V, are not coincident, as might be
expected as they both account for the same relative line strength.
Instead, the high-V group peaks at the same position as the main-V group,
while the low-V group peaks at the position of peak \nnh\ 1-0 emission
(DAM04).  The reason for this is not clear; perhaps it suggests that
non-LTE excitation anomalies, as seen in the 1-0 line (Caselli, Myers \&
Thaddeus 1995; Daniel, Cernicharo, \& Dubernet 2006) are present in the 3-2
line.  Modelling of the current observations with a non-LTE line code that
does not assume the hyperfines are in statistical equilibrium (Keto \&
Caselli 2010) could potentially address this question, but is beyond the
scope of this paper.  The satellite hyperfines are particularly strong in
N6, both in absolute intensity, and in their relative intensity compared to
the main-V group.  Their relative strength compared to the main-V group is
mostly due to the high total optical depth, as noted.  However, their
absolute intensity is about an order-of-magnitude brighter than has been
seen toward any other low-mass starless core, for example by comparison to
L1544, using just the 30-m data for each core (Caselli et al.\ 2002a,b;
Daniel et al.\ 2007).  

The map sizes as traced by the molecular line emission have been estimated
through two dimensional Gaussian fitting to the integrated maps, and
through approximate measurements by eye using the contour level tracing
50\% of the peak emission in the same maps.   The results are similar for
both methods for each line, suggesting the integrated emission can be
approximated by a Gaussian.  For \nnh\ 3-2, we only used the maps of the
low-V and high-V emission to estimate the size, so including \nnd\ we used
three maps in all.  All maps and methods give consistent results, with the
half-maximum diameter measured to be $\sim 3100 \times 1600$ AU, with
uncertainties of a few hundred AU for each axis.  The geometric mean
diameter is $\sim2200$ AU, which is smaller than the geometric mean
diameter of $\sim3400$ AU determined by DAM04 through \nnh\ 1-0
observations, likely due to the finer resolution of the observations
presented here (5\arcsec\ cf. 10\arcsec).  The ratio of major-to-minor axes
is about 2:1.  The core is well-resolved, as each axis is significantly
greater than the beam diameter ($640 \times 425$ AU for \nnd, $700 \times
460$ AU for \nnh).  

\subsection{Dust continuum emission}

\begin{figure}[!t]
\centering
\includegraphics[width=6in]{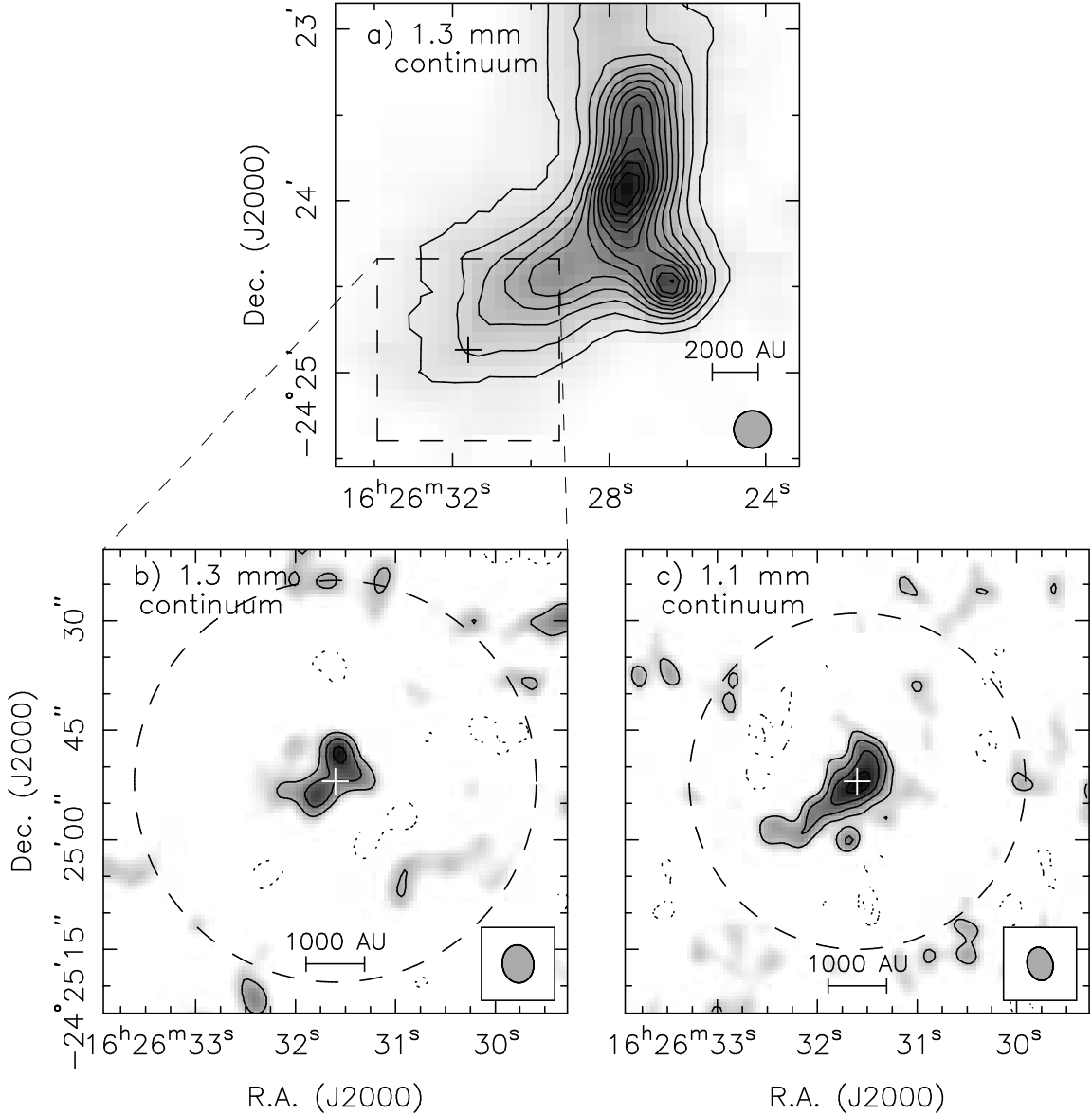}
\caption{Maps of (a) 1.3 mm emission (30-m), (b) 1.3 mm emission (SMA) and
(c) 1.1 mm emission (SMA).
Contours for (a) the 30-m observations are 10, 20, 30, ... times the $1\sigma$
sensitivity of 10 mJy/beam (Motte et al.\ 1998; DAM04).  
Contours for the SMA observations are 2, 3, 4, ... times the
$1\sigma$ sensitivity of (b) 2.6 mJy/beam (1.3 mm) and (c) 3.6 mJy/beam
(1.1 mm).  Dotted contours indicate negative levels.  The cross is the
position of the peak integrated \nnh\ 1-0 emission (DAM04).  The large
dashed circles indicates the size of the SMA primary beam.  The small grey
ovals at lower right each panel indicate the synthesised beam size.
%
}
\label{map-1p3mm}
\end{figure}

\begin{figure}[!t]
\centering
\includegraphics[height=3.5in,angle=-90]{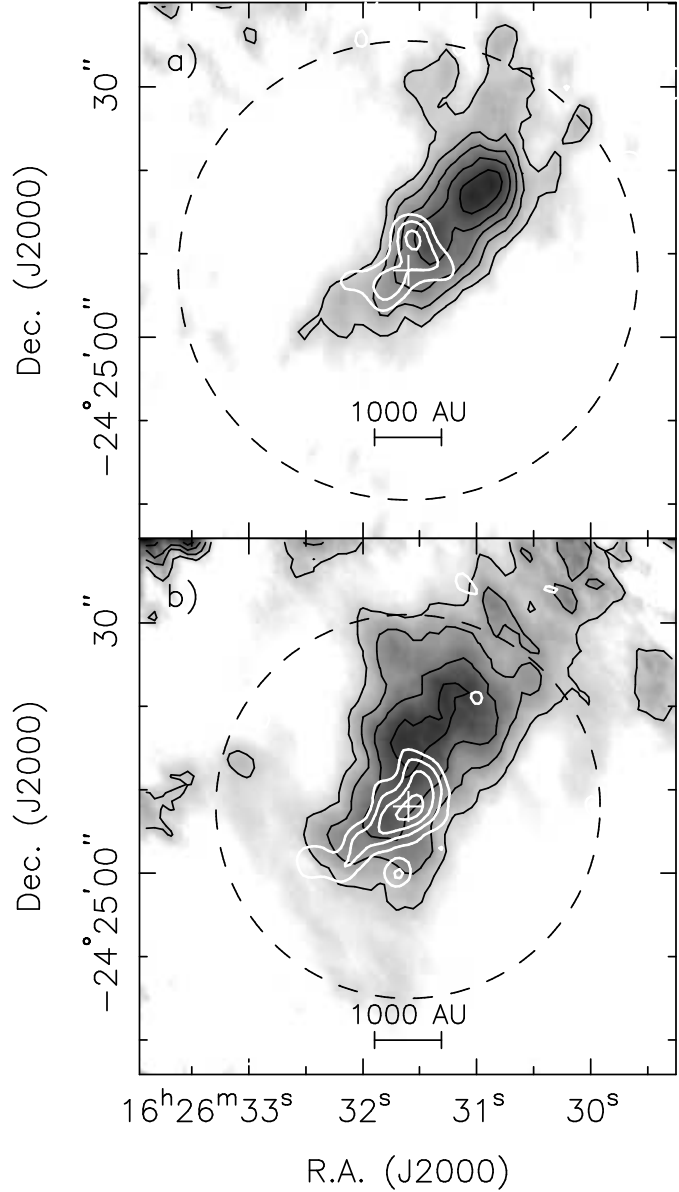}
\caption{Comparison of continuum and integrated line maps.  (a) Contours of
1.3 mm continuum emission (white; SMA) over contours of integrated \nnd\
3-2 emission (SMA+JCMT).  The greyscale is integrated \nnd\ 3-2 emission.
(b) Contours of 1.1 mm emission (SMA) over contours of integrated \nnh\ 3-2
emission (SMA+30-m).  Contour levels for \nnd\ are 10, 14, 18, ... times
the $1\sigma$ sensitivity of 0.12 Jy/beam \kms, while contour levels for
\nnh\ are 12, 16, 20, ... times $1\sigma$ sensitivity of 0.47 0.12 Jy/beam
\kms.  Continuum contour levels are the same as shown in
Figures~\ref{map-1p3mm}(a)--(b).  The white cross is the position
of the peak integrated \nnh\ 1-0 emission (DAM04).  The primary beam size
of the SMA for each observation is shown as the dashed circle. 
%
}
\label{map-compare}
\end{figure}

Weak dust continuum emission is detected at both 1.3 mm and 1.1 mm
(Fig.~\ref{map-1p3mm}).  Unlike the emission in the single-dish map
(Fig.~\ref{map-1p3mm}a; Motte et al.\ 1998), N6 is a local peak of emission
at these frequencies.  The interferometer has effectively filtered out the
larger scale bright emission to reveal the weak emission associated with
N6.  More extended emission is missing from the 1.3 mm map due to the array
configurations used, and this may be enough to cause the emission in this
map to look smaller than the 1.1 mm emission.  The dust emission is similar
in size and orientation at both wavelengths, with a small offset (3\farcs5)
between their peaks. However, this offset is at the 1$\sigma$ flux level
and so is probably not significant.  The orientation of the dust emission
is very similar to that of the molecular line emission
(Figure~\ref{map-compare}), and to that of the large scale dust emission.  

The mass of the region traced by the dust emission can be determined in the
standard manner (Hildebrand 1983) using the flux density of the region.
For these calculations we use the peak dust temperature of 20 K (Pon et al.\
2009), a gas-to-dust ratio of 100, and assume the dust opacity is
given by the commonly used ``OH5'' opacities that are believed to best
represent the dust properties within cold dense regions (Ossenkopf \&
Henning 1994; Evans et al.\ 2001).  Using the flux density above the 2
sigma level leads to a total mass estimate of 0.011 \Msun\ at both 1.3 mm
and 1.1 mm.  Using the 3 sigma level as the cutoff, the mass estimates are
0.006 \Msun\ at 1.3 mm and 0.005 \Msun\ at 1.1 mm.  Thus the values at 1.3
mm and 1.1 mm are in agreement.  At the 3 sigma intensity level the core
size is of order 1000 AU at both wavelengths, much smaller than the size of
the line emitting regions.  This small size may be due to large scale
structure being resolved out by the interferometer.

\subsection{Kinematic Structure}

Figures~\ref{fig-velmap} and \ref{fig-linemap} show the velocity and
linewidth maps of \nnd\ and \nnh\ 3-2.  The line velocity was determined
through fits of the hyperfine structure of each line using the hfs method
in CLASS\footnote{http://www.iram.fr/IRAMFR/GILDAS}, which performs a
simultaneous fitting of all hyperfine components using CERN's ``Function
Minimization and Error Analysis" package
MINUIT\footnote{http://wwwasdoc.web.cern.ch/wwwasdoc/minuit/minmain.html}.
MINUIT has been shown to produce accurate values for line velocities and
widths, even in the case of severe line overlap (Pon et al.\ 2009).
Typical uncertainties reported by CLASS for the fits reported here are
$\la$ 0.01 \kms\ in velocity and $\la$ 0.02 \kms\ in line-width, for \nnh,
and $\la$ 0.01 \kms\ in velocity and $\la$ 0.03 \kms\ in line-width, for
\nnd.

There is very little variation in line center velocity across the \nnd\ 3-2
map (Fig.~\ref{fig-velmap}a), with a possible hint of a gradient across the
long axis in the southern part of the core.  The variation in line center
velocity is mostly $<0.1$ \kms, larger than the typical uncertainty from
hyperfine fitting, and of the same order as the channel separation.  The
two integrated line peaks have velocities that are separated by about a
line width ($\sim0.25$ \kms).  The results are similar for \nnh\ 3-2, in
that there is very little variation in line center velocity across the map,
with the largest variation occurring along the western edge.  

DAM04 found that the linewidth of \nnh\ 1-0 over N6 is generally $\leq$ 0.3
\kms\ with a mean value of 0.25 \kms, with typical uncertainties of
0.005-0.01 \kms.  Our observations with higher angular resolution confirm
this result, and suggest that the linewidth varies by less than the channel
width of the observations over most of the map where significant emission
is present (Fig.~\ref{fig-linemap}).  There is a suggestion of large line
widths on the western edge, which may be due to the nearby dust continuum
source SM2, or simply due to lower S/N in the edges of the map.  There
appears to be an increase in \nnh\ line width of order $\sim0.03-0.04$
\kms\ near the \nnd\ SE peak, which is also near the position of the
continuum source (Fig.~\ref{fig-linemap}).  This increase is seen in the
\nnd\ data (Figure~\ref{spec-n2dp32}), as the line width of the SE peak of
\nnd\ ($0.279 \pm 0.018$ \kms) is significantly larger than that of the NW
peak ($0.217 \pm 0.015$ \kms) As in the case of \nnh\ 1-0 (DAM04), the
observed line width across N6 in the 3-2 lines of \nnh\ and \nnd\ is $\sim
0.25 \pm 0.02$ \kms.  Observations of the (1,1) and (2,2) lines of \nhhh\
indicate a gas temperature of $20 \pm 2$ K (Pon et al.\ 2009).  Using this
gas temperature, the thermal velocity dispersion $\sigma_{\rm T}$ is 0.26
\kms, implying that the non-thermal velocity dispersion $\sigma_{\rm NT}$
is $\sim 0.08$ \kms.  Thus the non-thermal motions within N6 are highly
subsonic, with $\sigma_{\rm NT} / \sigma_{\rm T} \sim 0.3$.

\section{Analysis}

\subsection{\nnh\ and \nnd\ Column Density}
\label{sec-coldens}

Line optical depths ($\tau$) and excitation temperatures (\tex) were
determined from fits to the hyperfine components of each transition, using
the fitting routines in CLASS (DAM04).  CLASS provides an estimate of the
optical depth, and the product [$J_\nu(T_{\rm ex})-J_\nu(T_{\rm
bg})$]$\tau$, where $J_\nu(T_{\rm ex})$ and $J_\nu(T_{\rm bg})$ are the
equivalent Rayleigh-Jeans excitation and background temperatures.  The
method of Caselli et al.\ (2002b; their appendix, in particular equation
(A4)) was used to determine the column density of \nnh\ and \nnd\ by
integrating over the hyperfine features.  This method assumes that the line
emission is optically thin.  Equation (A4) from Caselli et al.\ (2002b) is
repeated here, as it is important for the following discussion,

\begin{equation}
\label{eqn-coldens}
N_{\rm tot} = \frac{8 \pi W}{\lambda^3 A} \frac{g_l}{g_u}
\frac{1}{J_\nu(T_{\rm ex}) - J_\nu(T_{\rm bg})} \frac{1}{1 - \exp(-h\nu
/kT_{\rm ex})} \frac{Q_{\rm rot}}{g_l \exp(-E_l /kT_{\rm ex})} \,\, ,
\end{equation}

where $N_{\rm tot}$ is the total column density, $W$ is the integrated line
emission, $\lambda$ and $\nu$ are the wavelength and frequency of the
observations, $A$ is the Einstein coefficient, $g_l$ and $g_u$ are the
statistical weights of the lower and upper levels, and $Q_{\rm rot}$ is the
partition function.

Fits to the hyperfine structure of \nnd\ 3-2 show that this line is
optically thin at most positions (but with CLASS usually reporting values
$>$ 0.1), with the largest total opacities measured near to the map center
at $\tau \sim 2-3$.  Positions where the total optical depth was $\geq$ 1
were used to estimate a single excitation temperature for the whole map,
for which we find $\tex = 10.0 \pm 3.3$ K, so we assume a constant \tex\ of
10 K for \nnd.  For \nnd, the total integrated line emission and \tex\ were
used to calculate the column density $N$ at each map position.  The results
are shown in Figure~\ref{fig-coldens}(a).  The column density of \nnd\
ranges from $9.8 \times 10^{11}$ \cc\ to $4.7 \times 10^{12}$ \cc, with
values greater than $2 \times 10^{12}$ \cc\ over much of the map.

The \nnh\ 3-2 emission was found to be very optically thick over much of
N6, making it difficult to estimate $\tau$, and thus determine \tex.  In
addition, the saturated lines means that the observed integrated line
emission is only a lower limit of the true emission, and equation
(\ref{eqn-coldens}) is not valid for optically thick emission.  To overcome
this problem, a multistep approach was used to obtain an estimate of \tex
and measure the integrated line emission, so that the column density could
be determined.  Most of the optical depth of \nnh\ 3-2 is due to the main-V
hyperfine group.  The low-V and high-V groups only account for 0.0742 of
the total line strength (normalized to 1.0; Daniel et al.\ 2006; Pagani,
Daniel, \& Dubernet 2009).  The total integrated line emission was thus
found by integrating only over the low-V ($0.22-1.58$ \kms) and high-V
($5.07-6.18$ \kms) hf groups, and scaling by the inverse of their relative
line strength.  In the outer part of the \nnh\ map, the total optical depth
drops to reasonable values ($<15$), allowing the total line emission to be
measured, and compared to the value obtained using only the low-V and
high-V hf groups scaled by 1/0.0742.  At these positions, the results were
found to be in general agreement (better than 20\%).  Although the total
optical depth is high, the individual hyperfine features are optically thin
(39 hyperfine features in total, 17 in the low-V and high-V hyperfine
groups), and the total optical depth of the low-V and high-V hf groups
together are also thin in these data, or at most $\tau \sim 2$ with and
uncertainty of similar size.

While it was possible to obtain good fits to essentially every map position
using only the low-V and high-V hf groups, in most cases this resulted in
an optically thin fit ($\tau$ = 0.1 in CLASS), so that \tex\ is
unconstrained.  In order to obtain an estimate of \tex, full hf fitting is
needed.  Using only those positions away from the map center where the full
hf fit gives $\tau$ $<$ 20, we obtain $\tex = 10.0 \pm 2.2$ K.  
A full hf fit to a spectrum generated from the inner $8 \times 10$
positions, gives a similar result.  As we were unable to obtain a reliable
estimate for each individual map position, we assume that $\tex = 10 \pm 2$
K across the whole map.  This value is significantly lower than the value
of 17 K determined by DAM04 for \nnh\ 1-0, and the value of the kinetic
temperature of 20 K.  This difference could suggest that while the 1-0 line
is thermalized, the 3-2 line is not.  Alternatively, the denser interior of
the core, better traced by the 3-2 line, could be colder.  However, $\tex$
is fairly constant over the region mapped in \nnh\ 3-2, and the temperature
derived from dust observations is closer to 20 K, so this alternative is
the less likely of the two possibilities.

To determine the \nnh\ 3-2 column density, we assumed that the total column
density $N_{\rm tot} = N_{\rm hf}/0.0742$, where $N_{\rm hf}$ is the column
density of the outer hyperfines, calculated using the integrated line
emission of the low-V and high-V hyperfine groups, and assuming a constant
\tex\ of 10 K.  As shown in equation~(\ref{eqn-coldens}), the column
density $N$ (whether $N_{\rm tot}$ or $N_{\rm hf}$) is simply a
function of \tex, $f(\tex)$, times the integrated line intensity, $W$, so
that $N = W \times f(\tex)$.  The column density of \nnh\ determined in
this manner ranges from $3.5 \times 10^{12}$ \cc\ to $4.6 \times 10^{13}$
\cc, with most values being greater than $10^{13}$ \cc, and with a
significant fraction of the inner map region having values $> 2.5 \times
10^{13}$ \cc\ (Figure~\ref{fig-coldens}(b)).  The typical uncertainty in a
particular measurement of the column density is $N^{+100\%}_{-50\%}$.  
Similar values for the \nnh\ column density were found by
DAM04.

We have checked our results, using the outer hyperfine satellite groups and
assuming optically thin emission, against \nnh\ column densities determined
from hyperfine fits to the full hyperfine spectra (Caselli et al.\ 2002b;
Di Francesco et al.\ 2004; Friesen et al.\ 2010a).  We find that the
results are consistent, in that the values from the full fit are within the
uncertainties of the method we have used.  However, $N$ determined from the
full fit case are typically, but not systematically, higher (but are
sometimes lower) by up to 50\%.  Because the total
optical depth is so high its actual value is not well constrained by the
full fit at any particular position, so we prefer the method we have used
for estimating $N$.  

\subsection{Deuterium Fraction}

The ratio of \nnh\ and \nnd\ column densities can be used to estimate the
deuteration fraction within N6.  This is shown in Figure~\ref{fig-ratio},
where the ratio $N$(\nnd)/$N$(\nnh) is shown, compared to the integrated
intensity maps of each molecule.  From this Figure it can be seen that the
D/H ratio is of order 0.05 over a large fraction of the map, reaching
higher values toward the western side, of order 0.15.  These values are
larger than those determined by Pon et al.\ (2009), from lower resolution
observations.  Figure~\ref{fig-ratio} also shows that the NW \nnd\ peak has
a higher D/H ratio than the SE peak, as might be expected from Pon et al.\
(2009), where only the NW peak is clearly detected in the JCMT data.  This
result shows that Oph A-N6 has a high central degree of deuteration, and is
similar to values found for isolated low-mass starless cores (Crapsi et
al.\ 2005).  In some map locations the D/H value is close to the dividing
line of 0.1 used to characterize the isolated cores as prestellar or
starless, with the idea that prestellar cores are those closest to star
formation (Crapsi et al.\ 2005).  Of the prestellar cores identified by
Crapsi et al.\ (2005), all but one, like OphA-N6, have $N$(\nnh) $>
10^{13}$ cm$^{-2}$.
It is notable that even though the kinetic temperatures are near to 20 K,
where the D/H ratio should decrease dramatically (Caselli et al.\ 2008),
and significantly higher than in isolated cores, the D/H ratio is as high
as in most starless cores, if not higher.

\subsection{Structure \& Mass}
\label{sec-isoCylinder}

N6 is elongated and may represent a fragment of a filament.  The simplest
model of a filament is a self-gravitating isothermal cylinder, whose radial
density profile is (Ostriker 1964; Johnstone et al.\ 2003),

\begin{equation}
n(r) = \frac{n_0}{\left[ 1 + \left(\frac{r^2}{8H^2}\right) \right]^2} 
\, ,
\end{equation}

where $n_0$ is the peak number density, $r$ is the radial offset, and the
scale length $H$ is 

\begin{equation}
H^2 \equiv \frac{c^2}{4 \pi G \rho_0} \, ,
\end{equation}

where $c$ is the sound speed, $\rho_0$ the peak density, and $G$ is the
gravitational constant.

If N6 is viewed perpendicular to its axis, then the column density along
the line-of-sight is 

\begin{eqnarray}
N(r) &=& \frac{\pi}{2} \frac{n_0 H}{\left[ 1 +
\left(\frac{r^2}{8H^2}\right)^2 \right]^{3/2}} \\
&=& N_0 \frac{\pi}{4R} \frac{H}{\left[ 1 + \left(\frac{r^2}{8H^2}\right)^2
\right]^{3/2}}
\end{eqnarray}

where $N_0$ is the peak column density and $R$ is the radius.

\begin{figure}[!t] 
\centering
\includegraphics[width=6in,angle=270]{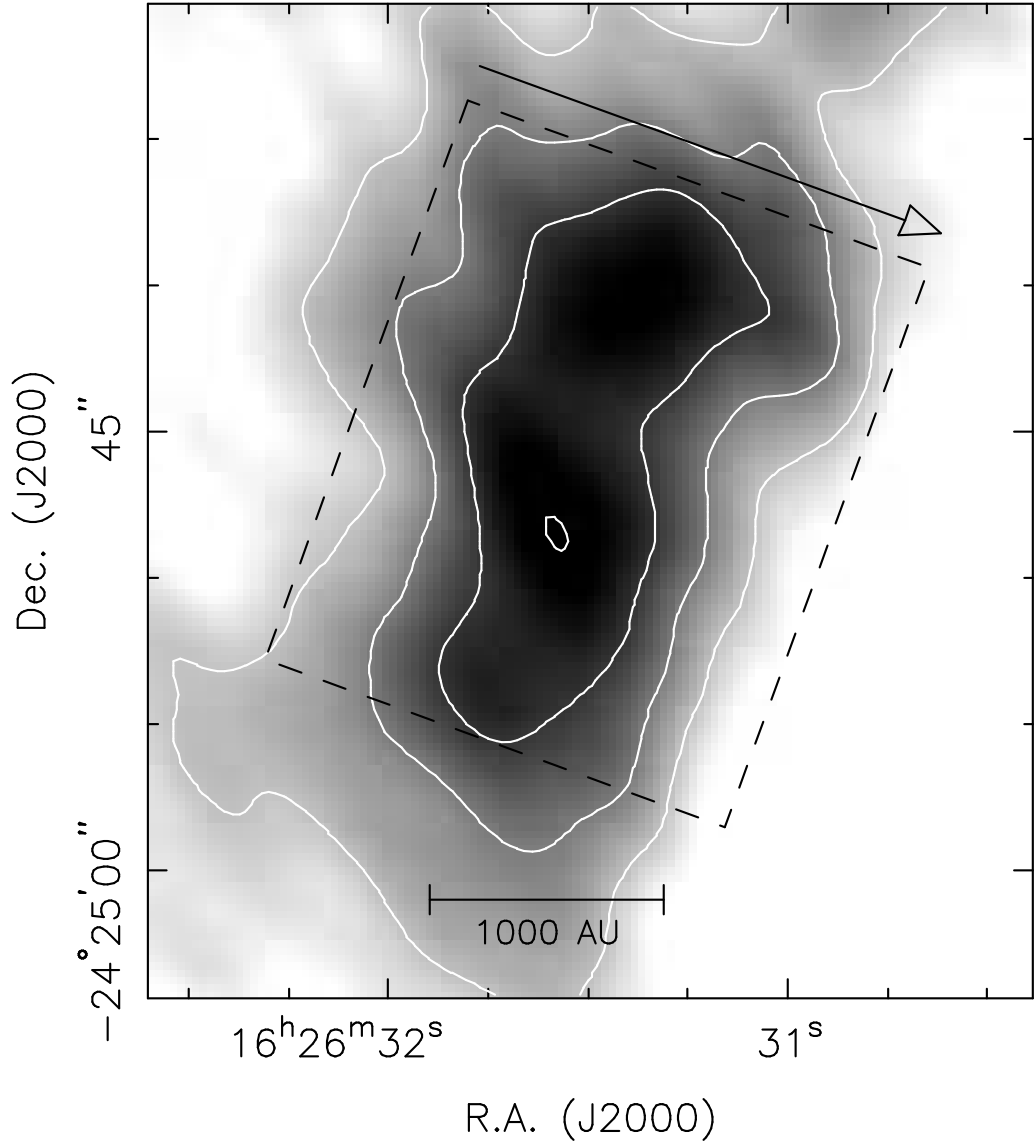}
\caption{
Location of the \nnh\ 3-2 radial column density cuts used to construct the
radial profile shown in Figure~\ref{fig-cylinder}.  The box indicates the
area from which the cuts were extracted, with the vector indicating the
direction of the cuts.  The image is the low resolution column density map,
with contour levels of 1.5, 2.5, 3.6, and 4.6 $\times 10^{13}$ \cc.
%
}
\label{fig-profile}
\end{figure}

\begin{figure}[!t] 
\centering
\includegraphics[width=6in,angle=270]{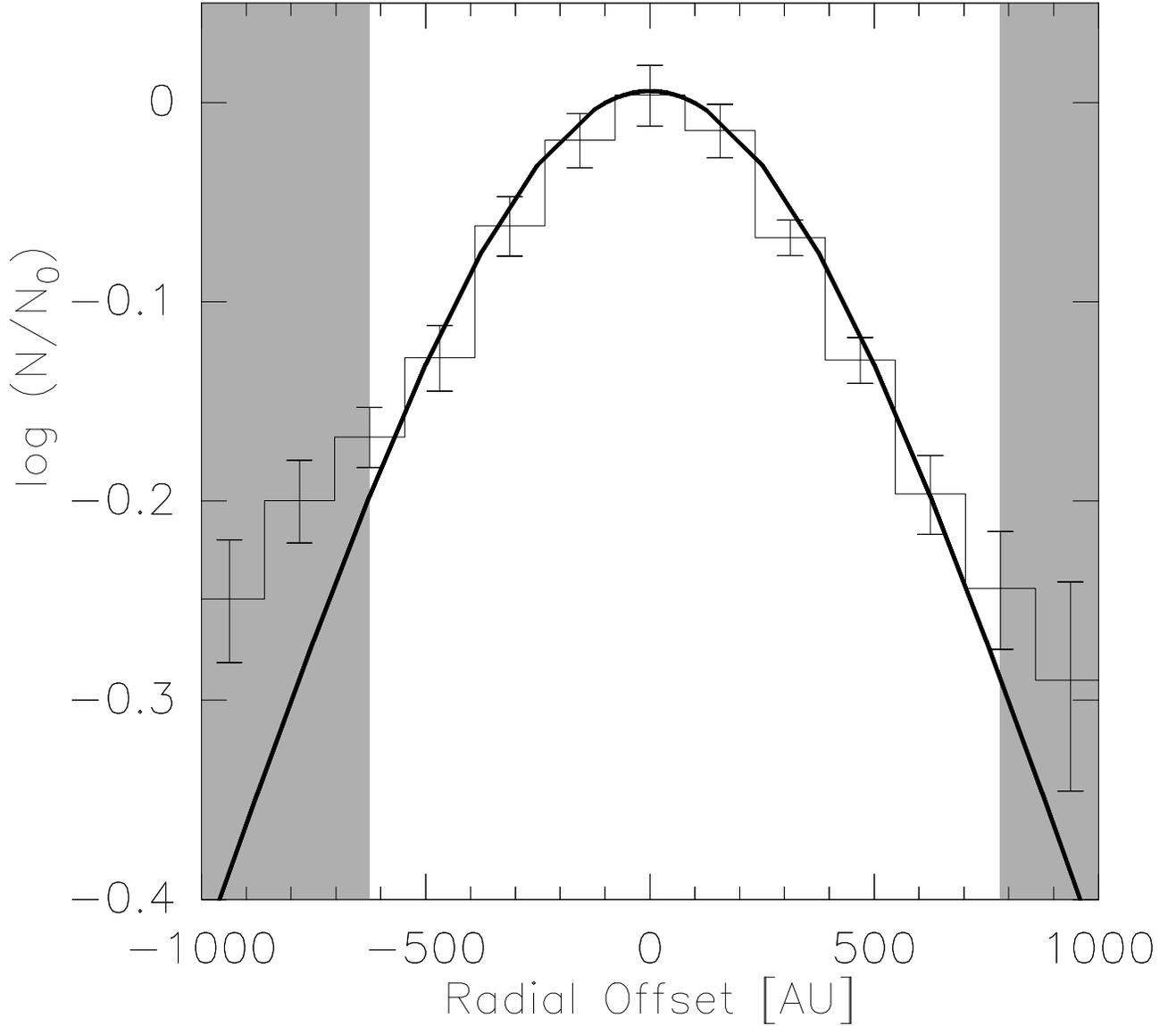}
\caption{
Normalized mean \nnh\ 3-2 radial column density cut parallel to the minor
axis of N6 (histogram), constructed from a series of cuts across N6 that
were averaged and then normalized to the peak value.  The dark continuous
line represents a model of an isothermal cylinder whose parameters are
described in \S\ref{sec-isoCylinder}.  The model shown here has the values
of radius (800 AU), kinetic temperature (20 K) and peak \nnh\ column
density ($4.6 \times 10^{13}$ cm$^{-2}$) fixed prior to comparison with the
data, an \nnh\ abundance ($2.75 \times 10^{-10}$), determined by matching
the model to the data (by eye), with a resultant  peak density of $7.0
\times 10^6$ cm$^{-3}$ and scale length of 365 AU.  The shaded grey areas
indicate where the data and model differ significantly and the core begins
to merge into the background emission.
%
}
\label{fig-cylinder}
\end{figure}

Figure~\ref{fig-cylinder} shows the radial column density profile across
the minor axis of N6 derived from \nnh\ 3-2 compared to the profile of an
isothermal cylinder (dark continuous curve).  This profile was constructed
from \nnh\ 3-2 data imaged with a 2\farcs4 beam and 1\farcs2 pixels
(Nyquist sampling), using the method of ``super-resolution" (Briggs 1994;
Chandler et al.\ 2005), in order to better sample the radial profile.
Eighteen independent, consecutive profiles were extracted across the major
axis at 1\farcs2 intervals along the major axis.  The region over which the
profiles were extracted is shown in Figure~\ref{fig-profile}.  Each profile
was normalized to its peak values, and the normalized profiles averaged
together to form the composite profile shown in Figure~\ref{fig-cylinder}.
This figure shows that the column density profile of N6 is very well
represented by an isothermal cylinder, as the model matches the data within
its 1$\sigma$ uncertainties at eight consecutive positions across the peal
of the profile.  The model allows the peak density and hence abundance of
\nnh\ to be estimated, keeping other parameters fixed at their previously
determined values; radius $R$ = 800 AU, temperature of 20 K (Pon et al.\
2009), and peak \nnh\ column density of 4.6 $\times 10^{13}$ cm$^{-2}$.
Using these values, we find a good match to the data, as shown in
Figure~\ref{fig-cylinder}, assuming a constant \nnh\ abundance $X_{N_2H^+}
= 2.7\pm0.2 \times 10^{-10}$, resulting in values of peak density $n_0 =
7.1^{+0.6}_{-0.5} \times 10^6$ cm$^{-3}$, and scale length $H =
362^{+12}_{-14}$ AU.  Allowing for a 5 pc uncertainty in the distance does
not change these values.

Even though N6 is not a local dust emission peak, DAM04 estimated the
column density, $N$(H$_2$), to be $3 \times 10^{23}$ cm$^{-2}$ using the
dust continuum emission, assuming isothermal dust at a temperature of 20 K.
From this and their value for $N$(\nnh) they infer an abundance $X$(\nnh)
of $3 \times 10^{-10}$, in very close agreement with the value used here
that provides an excellent match between the isothermal cylinder model and
the data.  This abundance is in good agreement with values inferred for
isolated low-mass cores, including the evolved prestellar cores discussed
above (Benson, Caselli \& Myers 1998; Caselli et al.\ 2002c; Crapsi et al.\
2005).  

The mass per unit length of an isothermal cylinder is 

\begin{equation}
M(r) = 2 \pi \rho_0 \int_{0}^{R} r dr 
\left[ 1 + \left(\frac{r^2}{8H^2}\right) \right]^{-2}
\end{equation}

After integrating, the mass of a cylinder of length $L$ can be written:

\begin{equation}
M = L \, \frac{2c^2}{G} \left[1 + \left(\frac{2c^2}{\pi G \rho_0
R^2}\right)\right]^{-1}  \, .
\end{equation}

For $T = 20$ K, $n_0 = 7.1 \times 10^6$ \cm{-2}, $R$ = 800 AU, and $L$ =
3100 AU, the mass is $M = 0.18\pm0.02$ \Msun, where the uncertainty is due
to the uncertainties in $n_0$ given above and the distance uncertainty.  

We can determine the total mass traced by \nnh, using the \nnh\ column
density map (Fig.~\ref{fig-coldens}(b)), with the result for the \nnh\
abundance.  The map gives the column density per pixel, from which the mass
per pixel ($M_p$) can be determined, and hence the total mass, using

\begin{equation}
\label{eqn-mass1}
M_p = X \, \mu_m \, A_p \, N_X \,\, ,
\label{eqn-masspix}
\end{equation}

where $\mu_m$ is the mean particle mass (2.37 amu; Stahler \& Palla 2005;
Kauffmann et al.\ 2008), $A_p$ is the area per pixel, $X$ is the abundance
of the molecule used, and $N_X$ is its column density.  In a Nyquist
sampled map, the total mass is then just the sum over all pixels.  For
\tex\ = 10 K and $X_{N_2H^+}$ of $2.7 \times 10^{-10}$, we measure $M =
0.29^{+0.05}_{-0.04}$ \Msun\ for positions within the half-power level of
the column density map.  The uncertainties come from the uncertainties in
$X$ and the distance.  The change in mass by assuming \tex\ = 9 or 11 K is
much smaller than either of these. 

The critical mass is the mass of a condensation whose radius is equal to
the shortest wavelength of a periodic perturbation that will grow.  Larson
(1985) has studied the critical mass for fragmentation of a number of
geometries, and for an isothermal filament (i.e., a cylinder) finds (Larson
1985, equation 21)

\begin{eqnarray}
M_c &=& \frac{7.88 c^4}{G^2 \mu_m N} \\
&=& 1.1 \left(\frac{T}{20\,{\rm K}}\right)^2 \left(\frac{10^{23}\,\,
\cm{-2}}{N}\right) [\Msun].
\end{eqnarray}

With $T=20$ K and $N = 1.7 \times 10^{23} \cm{-2}$ (from the peak \nnh\
column density and $X$), $M_c = 0.63^{+0.05}_{-0.04}$ \Msun.  This value is
within about a factor of 2 of the mass computed for N6 of 0.29 \Msun\ from
eqn.~(\ref{eqn-masspix}).  Given the uncertainties in computing masses,
such as determining the ``size" of a core, and our method of measuring $N$
at each position, this result suggests that N6 is consistent with having
formed from the fragmentation of an isothermal filament, in this case Oph
A.  

\section{Discussion}

\subsection{Kinematics}

Internally, Oph A-N6 is rather quiescent.  It shows very narrow \nnh\ and
\nnd\ line-widths of about 0.25 \kms\ across its extent, barely more than
the thermal line width for the measured gas temperature of 20 K, of 0.18
\kms.  Its non-thermal motions are very sub-sonic, but the surrounding gas
shows significantly larger line-widths (DAM04; Andr\'e et al.\ 2007; Pon et
al.\ 2009) suggesting that N6 has lost any turbulent motions it may have
had.  The lack of significant variation in line centroid velocity and
line-width over the core indicate that N6 is an example of a coherent core,
as has been seen in more isolated cores (Barranco \& Goodman 1998; Goodman
et al.\ 1998; Caselli et al.\ 2002a; Tafalla et al.\ 2004; Pineda et al.\
2010).  This result suggests that small non-thermal motions typical of
isolated cores are found in some starless cores within turbulent molecular
clouds.

Observations of HCO$^+$ and DCO$^+$ 3-2 show the expected signature of
inward motions (Evans 1999; Pon et al.\ 2009), but the complex hyperfine
structure of \nnh\ and \nnd\ 3-2 makes identifying any similar signature in
these lines impossible.  In addition, the very narrow line widths of \nnh\
1-0 together with the spectral resolution and signal-to-noise of the data
make it difficult to identify any signature of inward motions (DAM04),
regardless of the hyperfine structure.  Data with finer spectral resolution
and improved signal-to-noise are required to search for inward motions in
\nnh.  However, the very narrow line-widths already suggest that any inward
motions on the size scales probed by \nnh\ ($\sim300$ AU) must be small.
The ratio of non-thermal-to-thermal line-width in N6 is about 0.3, which is
lower than observed in most starless dense cores in Perseus (Walsh et al.\
2007; H.~Kirk et al.\ 2007; Rosolowsky et al.\ 2008), for dense cores
elsewhere in Ophiuchus (Andr\'e et al.\ 2007; Friesen et al.\ 2009), or for
most isolated low-mass dense cores (Myers 1983).  Further, the absence of
line-broadening toward the center of N6 suggests a lack of a central
source.  The motions observed in HCO$^+$ may be infall onto the core,
rather than core collapse (Pon et al.\ 2009).

\subsection{Dust Emission}

Starless cores are usually defined through observations of the dust
continuum or molecular lines in single-dish observations at millimeter
wavelengths, with angular resolution 10-20\arcsec\ (typically the line
observations are of lower resolution than the continuum).  As a result, by
definition they generally only show a single peak of emission, and fairly
simple structures, being round or elongated with small aspect ratios (less
than 2).  When observed with an interferometer, which acts as a spatial
filter, many such cores are not detected, or still only appear as single
peaks of emission, due to their smooth large-scale structure and lack of
significant sub-structure (Williams \& Myers 1999; Williams et al.\ 1999,
2006; Harvey et al.\ 2003a; Olmi et al.\ 2005; Schnee et al.\ 2010).
Combining the single-dish and interferometer line data, as we have done
here, allows the small scale structure to be studied, without concerns
about missing flux.  These studies usually show that starless cores do not
break up into sub-cores on small scales.   One exception is L183, which is
composed of 3 sub-cores in \nnh\ 1-0 (J.~Kirk et al.\ 2009), but it shows a
very elongated structure in single dish maps, so perhaps this is not too
surprising.   

The nature of the compact dust emission detected toward the peak of
integrated \nnh\ emission is unclear, given that N6 is not a local maximum
in single-dish continuum observations between 1300 and 450 \micron, with
10-15\arcsec\ resolution (Motte et al.\ 1998; Wilson et al.\ 1999;
Johnstone et al.\ 2000).  However, a dust temperature map derived from the
ratio of 450-to-850 \micron\ flux, assuming a constant dust emissivity,
shows a similar structure to the \nnh\ maps, although with lower resolution
(Pon et al.\ 2009).  The dust temperature map shows a peak of 20 K at the
\nnh\ peak, and is elongated in the NW-SE direction.  It is not yet known
if the gas temperature varies in a similar manner on similar scales, as the
\nhhh\ observations only have a resolution of about 30\arcsec.  However,
\nhhh\ may not probe the highest densities toward the center of N6, and so
determining the gas temperature there with confidence will be difficult.
Supporting evidence for a relatively constant gas temperature within N6
comes from the comparison of its column density profile with that of an
isothermal cylinder (Fig.~\ref{fig-cylinder}), and from the almost
constant line-widths.

N6 is embedded within the Oph A ridge, and the large column of dust due to
the ridge may make it difficult to distinguish a compact core within it as
a separate entity.  The dust temperature map suggests that it is a local
temperature maximum, at about 20 K (Pon et al.\ 2009).  This result is
unlike those in detailed studies of isolated starless cores, which show
flat temperature profiles in low-resolution observations (Jijina et al.\
1999; Tafalla et al.\ 2004), but a drop in temperature toward the core
center in observations with finer resolution (Crapsi et al. 2007).  The
mass of dust seen in N6 is very low, only of order 0.005-0.01 \Msun, and
the inferred peak column density of $\sim 1.3 \times 10^{22}$ \cm{-2}\ is
an order of magnitude below that found from \nnh\ observations, and from
single-dish continuum observations.

Recently, interferometers have detected compact millimeter dust emission
toward three ``starless" cores (Chen et al.\ 2010; Enoch et al.\ 2010;
Pineda et al.\ 2011; Dunham et al.\ 2011).  Supporting evidence, in the
form of CO outflows or faint, compact 70 \micron\ emission, and SED
modeling, suggests that in all cases the emission is due to an internal
heating source of very low temperature ($>$100 K), and the inferred
luminosities are very low ($<$0.1 \Lsun).  These cores are all candidates
to be the long theorized first hydrostatic core (FHSC; Larson 1969; Boss \&
Yorke 1995; Omukai 2007; Tomida et al.\ 2010), although none are in
complete agreement with theoretical predictions.  They have outflows that
are too fast (L1448-IRS2E -- Chen et al.\ 2010), too collimated
((L1448-IRS2E -- Chen et al.\ 2010; Per-Bolo 58 -- Dunham et al.\ 2011),
or are detected at too short a wavelength (Per-Bolo 58 -- Enoch et al.\ 2011),
to be consistent with current models.  For one source, L1451-mm, the
observations are in better agreement with theoretical models, but a model
of a protostar plus a disk provides an equally good fit to its SED and
continuum interferometric visibilities (Pineda et al.\ 2011).
Theoretically, a FHSC is expected to be of low mass, with a maximum value
of order 0.05 \Msun, essentially undetectable at $<100$ \micron, and with
an observed SED resembling a blackbody at 30 K (Omukai 2007; Saigo \&
Tomisaka 2011).  A low-velocity, compact outflow may also be present.  Due
to crowding, it is difficult to determine if 70 \micron\ Spitzer emission
is present toward N6, and as stated it does not show a local maximum in
450-1300 \micron\ emission in single-dish observations.  Observations to
search for the presence of an outflow in CO 2-1 are compromised by the
bright outflow from VLA 1623 (Andr\'e et al.\ 1990), that passes close to
N6 and is detected in the SMA observations presented here.  The compact 1
mm dust emission detected with the SMA has the right mass to be considered
a candidate FHSC, but further evidence is needed, particularly detections
at other wavelengths.  At present there is insufficient information to
suggest whether N6 is a better FHSC candidate than the other candidates.

\subsection{Comparison to other Starless Cores}

Studies of starless cores with resolution similar to the one presented here
are rare, particularly in molecular lines.  This is a greater problem for
cores in clusters, which are typically smaller than more isolated cores,
such as those in Taurus (Ward-Thompson et al.\ 2007, and references
therein).  Studies with 10-15\arcsec\ resolution have been made for the
cluster-forming regions in Ophiuchus (125 pc; Motte et al.\ 1998; Johnstone
et al.\ 2000; DAM04; Simpson et al.\ 2008; Friesen et al.\ 2009), Perseus
($\sim235$ pc; Hatchell et al.\ 2005; Walsh et al.\ 2007) and Serpens
($>$300 pc; Testi \& Sargent 1998; Williams \& Myers 1999), and for
isolated cores in Taurus (140 pc; Ward-Thompson et al.\ 1994; Caselli et
al.\ 2002a,b; Tafalla et al.\ 2002; J.~Kirk et al.\ 2005).  A comparison of
the properties of N6 with the cores in these studies shows that N6 is
denser than most starless cores, by about an order-of-magnitude ($10^7$
\cube\ cf. $10^6$ \cube).  These observations typically do not have the
resolution of our observations of N6, and derived average densities of
$10^6$ \cube\ over the central 1000 AU could be consistent with peak
densities of $\sim 10^7$ \cube\ (Keto \& Caselli 2010).

N6 is smaller than most cores in all three cluster-forming regions listed
above, (Walsh et al.\ 2007; Friesen et al.\ 2009), but this could be partly
an effect of resolution, as this study has finer resolution by at least a
factor of two over previous work.  The small size could also be partly due
to the molecular transition used, as we have used higher J transitions that
preferentially trace higher column density material.  Walsh et al.\ (2007)
list a few cores with sizes comparable to N6, but these are at the limit of
their resolution.  N6 is significantly smaller than all isolated cores that
have been well resolved, whether studied in line-emission, dust continuum,
or extinction (Ward-Thompson et al.\ 1999; Bacmann et al.\ 2000; Crapsi et
al.\ 2005; Kandori et al.\ 2005; Kauffmann et al.\ 2008).

As noted earlier, the \nnh\ linewidths in N6 are narrower than almost all
other cluster cores (Andr\'e et al.\ 2007; Walsh et al.\ 2007; H.~Kirk et
al.\ 2007; Friesen et al.\ 2010a), and are almost totally due to thermal
motions (as discussed in detail in DAM04).  The linewidth barely varies
across N6, and it is an excellent example of a coherent core, a core where
the non-thermal motions are subsonic, and constant, so that it appears to
be cut-off from the surrounding turbulent gas (Mouschovias 1991; Myers
1998; Barranco \& Goodman 1998; Goodman et al.\ 1998; Caselli et al.\
2002c; Pineda et al.\ 2010).

The \nnh\ column density in N6 is larger than most cores elsewhere in
Ophiuchus (Friesen et al.\ 2010a) and Perseus (H.~Kirk et al.\ 2007), with
a peak value of $\sim5 \times 10^{13}$ \cc\ compared to values of $\sim
10^{13}$ \cc.  This peak value is about three times greater than the peak
value observed in a sample of 28 isolated starless cores, and about eight
times greater than the sample mean ($\sim8 \times 10^{12}$ \cc; Crapsi et
al.\ 2005; see also Daniel et al.\ 2007).  Similarly, the \nnd\ column
density of N6, $\sim5 \times 10^{12}$ \cc, is greater than that of cores in
Oph B, where the peak value is $\sim7 \times 10^{11}$ \cc, and greater than
the mean value of 25 isolated starless cores, of $<10^{12}$ \cc\ (Friesen
et al.\ 2010b; Crapsi et al.\ 2005; see also Daniel et al.\ 2007).
However, the peak value of N(\nnd) for isolated starless cores, observed
toward L1544, L429 and L694-2, is similar to N6 (Crapsi et al.\ 2005).  

The deuterium fraction, ranging from a mean near 0.05 to a maximum value of
about 0.15, is similar to that seen in 28 isolated cores (range 0.03 --
0.44), where 22 of the cores have D/H $<$ 0.1 (Crapsi et al.\ 2005).  The
range of values in N6 is also similar to that of the cluster-core Oph B2,
which has a peak of 0.16 but with most of the core showing values around
0.03 (Friesen et al.\ 2010b).  The mean temperature of the isolated cores
is about 10 K, while the mean in Oph B2 is higher at around 13-14 K.  The
deuterium fraction is expected to be significantly higher in the cold
($<$20 K) dense interiors of starless cores than the cosmic D/H ratio.
This is due to two main factors.  First, the main pathway for the formation
of H$_2$D$^+$, the parent molecule of deuterated molecules, is exothermic
by 230 K, and the backward reaction is not available.  Second, CO will
freeze out at temperatures below 20 K, removing the main destroyer of
H$_2$D$^+$.  However, N$_2$ should also freeze out at essentially the same
rate at CO at $<$20 K, so the picture is not so simple.  A number of more
subtle factors that affect the D/H ratio, such as grain size (surface
chemistry), ionization rate, ortho-to-para H$_2$ value, and the CO
depletion factor are examined in detail by Caselli et al.\ (2008).  They
show that the D/H ratio can still be relatively high near 20 K, but
drops sharply at $<$15 K.  
One obvious explanation for the relatively high values of D/H in N6 may be
temperatures a few degrees lower than 20 K, but the complete reason for the
high D/H values is likely to be more complicated.  The D/H ratio is largest
away from the dust temperature peak, to the NW, and Pon et al.\ (2009) do
infer a radial temperature drop.  

Thus, N6 appears to be denser and smaller than starless cores in both
cluster-forming and isolated environments.  While a very small number of
cluster-cores have a similar size, no isolated core does, and no starless
core has a mean density as high.  

\subsection{Structure and Evolution}

Starless cores have typically been modeled with spherically symmetric
geometries, with a radial density profile that is almost constant at small
radii but decrease as a power law at larger radii (Ward-Thompson et al.\
1994, 1999; Bacmann et al.\ 2000; Evans et al.\ 2001; Kandori et al.\ 2005;
J.~Kirk et al.\ 2005).  However, N6 is clearly elongated, with an aspect ratio
of at least 2:1, as observed in many cores (Myers et al.\ 1991; Ryden
1996), suggesting that it is prolate or filamentary in nature.  Over its
half-maximum size its dust temperature is fairly uniform at about 20 K, to
within 1-2 K, decreasing at larger distances (Pon et al.\ 2009).  We have
thus compared the column density profile of N6 to that of an isothermal
cylinder (Ostriker 1964; Curry 2000), finding an extremely good match
between the data and the model  (Fig.~\ref{fig-cylinder}).  

The mass of the isothermal model cylinder is $\sim0.2$ \Msun, similar to
the observationally derived mass of 0.3 \Msun, given the uncertainties in
the input values (cylinder size, column density per pixel, temperature).
Similarly, the critical mass for fragmentation of an isothermal filament
with properties similar to those of N6 is $\sim0.6$ \Msun\ (Larson 1985),
also close to the observational value given the uncertainties (in
particular as the observed value of mass depends on column density, whereas
the critical cylinder mass depends on its inverse).  The
excellent match between the column density profile of N6 and the isothermal
cylinder model, and the similarity of the observed mass to the mass of an
isothermal filament with the properties of N6, strongly suggest that N6
has formed via fragmentation of the Oph A filament, and is in a critical
state at the beginning of star formation, or has already started the
star-formation process (as evidenced by the low-mass compact dust continuum
emission).  N6 is aligned with its parent core, Oph A, supporting the view
that it has formed as the result of fragmentation of Oph A along its axis.  

Although non-spherically symmetric models are rarely used, they have been
quite successful in explaining the observed elongated structure of dense
cores and filamentary nature of molecular clouds (Harvey et al.\ 2003b;
Johnstone et al.\ 2003).  It has also been shown that the density profiles
of collapsing centrally condensed (``Bonner-Ebert") spheres and cylinders
are remarkably similar, and distinguishing between these cases
observationally, using only column density maps, may not be possible (Myers
2005).  Resorting to the ease of using spherical models without considering
alternatives should be avoided, a message that is often given without heed
(Hartmann 2004).  

Harvey et al.\ (2003b) performed a detailed study of the isolated starless
core L694-2, through near-infrared extinction mapping.  They found that
spherical models where the radial density profile is described by a
power-law, or a Bonnor-Ebert sphere, did not provide accurate matches to
the data.  Instead they found that a cylindrical model, like that
described here, provided an excellent fit to their data. In the case of
L694-2, the cylinder is inclined to the line-of-sight, so initially it was
not obvious that a cylindrical model was needed.  The steep power-law index
that best matched the data, significantly steeper than predicted by
inside-out collapse, motivated the cylindrical model.  

Many distant infrared dark clouds and nearby star-forming complexes have
filamentary or elongated structures (Schneider \& Elmegreen 1979; Myers
2009), and in at least one case, that of the infrared dark cloud
G11.11-0.12, the radial profile of 850 \micron\ emission closely matches
that of an isothermal filament of radius $\approx$ 0.1 pc over a length of
more than 10 pc (Johnstone et al.\ 2003).   While many molecular clouds
have been compared to models of infinite or finite sheets, models of
individual clouds as cylinders are almost absent, although theoretical
studies suggest that such models could provide good matches to available
data (Curry 2000).  New results from the Herschel Space Observatory show
filamentary structures in nearby molecular clouds (Andr\'e et al.\ 2010;
Arzoumanian et al.\ 2011), and modelling suggests radial profiles much
shallower than isothermal cylinders, at radii $>$ 0.1 pc.  At such large
radii the assumptions of constant temperature and/or hydrostatic
equilibrium are unlikely, so this result is not surprising.  The
observations do not have the resolution to probe scales similar to that of
N6.  The profiles could be consistent with collapsing polytropic cylinders
(Arzoumanian et al.\ 2011), or magnetized filaments in virial equilibrium
(Fiege \& Pudritz 2000).  It will be interesting to observe cores within
these filaments with finer resolution to study their structure.

The physical, kinematic, and chemical properties of dense cores have been
used to assess their evolutionary state (Ward-Thompson et al.\ 1999; Crapsi
et al.\ 2005; Di Francesco et al.\ 2007).  Crapsi et al. (2005) searched
for evolutionary indicators in a sample of 31 isolated cores, using
observations of \nnh, \nnd, CO and the dust continuum.  They proposed eight
chemical and kinematic evolutionary indicators, and identified as
``evolved" those cores that met at least four of the conditions.  These
conditions include large column densities of \nnh\ and \nnd, a large
deuterium fraction, large CO depletion and central density, broad
linewidths, infall asymmetry in the line profiles, and compact central
regions.  All of these conditions, with the exception of CO depletion, can
be tested in N6 with our data.  

The peak column densities of \nnh\ ($4.6 \times 10^{13}$ \cm{-2}) and \nnd\
($4.7 \times 10^{12}$ \cm{-2}) are greater than the dividing values given
by Crapsi et al. (2005), of $8.5 \times 10^{12}$ \cm{-2}\ and $1.0 \times
10^{12}$ \cm{-2}, respectively.  Similarly, the peak of the deuterium
fraction, 0.15, is greater than the value of 0.1 used by Crapsi et al. to
separate candidate prestellar cores from starless cores.  The peak density
we determine, $7.0 \times 10^6$ \cm{-3}, is far greater than the separator
of $5.1 \times 10^5$ \cm{-3}\ used by Crapsi et al., as is the degree of
central concentration (this is not a function of resolution, as the single
dish data have been included in our results).  The narrow line widths, and
hyperfine structure, makes the search for line skewness difficult.  As
mentioned, Pon et al.\ (2009) observe infall asymmetry in HCO$^+$, so there
is some evidence of contraction in N6.  However, the \nnh\ and \nnd\ lines
widths are of similar size to the separating value of 0.25 \kms\ used by
Crapsi et al.\ (2005).   So of the seven evolutionary indicators we can
measure, five clearly indicate that N6 is an evolved prestellar core,
according to the analysis of Crapsi et al.\ (2005), while the other two
indicators are borderline (line widths) or unclear (infall asymmetry or
line skewness).  These results strongly suggest that N6 is at an advanced
stage of prestellar evolution.  The compact dust continuum emission further
supports this idea.

\section{Summary \& Conclusions}

We have presented high spatial ($\sim$500 AU) and spectral (0.07 \kms)
resolution observations in \nnh\ 3-2, \nnd\ 3-2, and the dust continuum at
1-mm, of the starless core Oph A-N6, embedded within the Oph A molecular
ridge in the Ophuichus cluster-forming molecular cloud.  These are the
highest resolution observations of a starless dense core presented to date.
Such a small condensation as N6 would be hard to detect with coarser
angular resolution observations, especially in clouds much more distant
than Ophiuchus.  

The major results of this study are summarized below:

\begin{enumerate}

\item The observations reveal a compact dust continuum peak, of size
$\sim$1000 AU and mass 0.005-0.01 \Msun, not seen in single dish
observations, except in a dust temperature map (Pon et al.\ 2009).   The
small size and mass suggests it might be the first indication of collapse,
either representing a temperature increase, or possibly a first hydrostatic
core.  

\item The size of N6 from the line observations is larger, with a projected
half-power diameter of $3100 \times 1600$ AU, and thus an aspect ratio of
2:1.  The \nnh\ and \nnd\ integrated line maps show slightly different 
position angles, probably representing chemical variations, or the very
high optical depth in \nnh.

\item Very little variation is seen in either linewidth or line center
velocity in either line across their maps.  The variations are so small
that N6 appears to be a coherent core, with very small non-thermal motions.

\item The peak column densities are $4.6 \times 10^{13}$ \cc\ for \nnh,
and $4.7 \times 10^{12}$ \cc\ for \nnd.  The positions of peak column
density are offset, with the \nnd\ peak located to the NW of the \nnh\
peak.

\item The deuterium fraction has a peak value of 0.15, and is
greater than or about equal to 0.05 over much of the mapped area.  The
maximum value of deuteration lies in the NW, and not at the position of the
dust continuum peak, nor the \nnh\ peak.

\item The column density profile of N6 across its minor axis, as determined
from the \nnh\ observations, is very well represented by an isothermal
cylinder (at 20 K), of peak density $7.1 \times 10^6$ \cm{-3}, and \nnh\
abundance $2.7 \times 10^{-10}$.  

\item The mass of N6, determined from the mapped positions, lies in the
range 0.25-0.34 \Msun, depending strongly on the assumed \nnh\ abundance
($2.7 \pm 0.2 \times 10^{-10}$).  The low value compares favourably to the
mass determined from the cylindrical analysis, of $\sim0.2$ \Msun, while
the high value compares well to the critical mass for fragmentation of an
isothermal filament with similar properties, of $\sim0.6$ \Msun.  

\item Compared to isolated low-mass cores, Oph A-N6 shows similar narrow
line widths and small velocity variation, with a deuterium fraction that is
similar to ``evolved" dense cores.  It is significantly smaller than
isolated cores, with larger peak column density and volume density, while
the previously measured kinetic temperature is significantly higher than
isolated starless cores.

\end{enumerate}

These results strongly suggest Oph A-N6 has formed from the fragmentation
of the Oph A filament, and is a precursor to a low-mass star in a
cluster-forming region.  The results also suggest Oph A-N6 has completed
almost all of its prestellar evolution, and may even have begun to form a
star.  

\acknowledgments

This research is supported in part by the National Science Foundation under
grant number 0708158 (T.L.B.).  We thank Andy Pon and Rachel Friesen for
sharing results in advance of publication, and Rachel Friesen and Chris de
Vries for checking column density estimates for optically thick \nnh.  We
thank Mark Gurwell for his diligent maintenance of the ``Submillimeter
Calibrator List".  The Submillimeter Array is a joint project between the
Smithsonian Astrophysical Observatory and the Academia Sinica Institute of
Astronomy and Astrophysics and is funded by the Smithsonian Institution and
the Academia Sinica.  The James Clerk Maxwell Telescope is operated by The
Joint Astronomy Centre on behalf of the Science and Technology Facilities
Council of the United Kingdom, the Netherlands Organisation for Scientific
Research, and the National Research Council of Canada.  Based on
observations carried out with the IRAM 30\,m telescope. IRAM is supported
by INSU/CNRS (France), MPG (Germany) and IGN (Spain).  This research has
made use of NASA's Astrophysics Data System Bibliographic Services


\clearpage
\begin{figure}[!t]
\centering
\includegraphics[width=6in]{f1.eps}
\caption{Integrated combined single-dish + interferometer line maps of (a)
\nnh\ 1-0 (from DAM04), (b) \nnd\ 3-2, and (c) \nnh\ 3-2.  The contour
levels are (a) 6, 9, 12 ... times the $1\sigma$ sensitivity of 0.6 Jy/beam
\kms\ for \nnh\ 1-0, (b) 9, 12, 15 ... times the $1\sigma$ sensitivity of 0.12
Jy/beam \kms\ for \nnd\ 3-2, and (c) 12, 15, 18 ... times the $1\sigma$
sensitivity of 0.47 Jy/beam \kms\ for \nnh\ 3-2.  The white cross is the
position of the peak integrated \nnh\ 1-0 emission from DAM04.  The large
dashed circles in (b) and (c) indicate the size of the SMA primary beam
(full-width at half-maximum sensitivity).
The small grey ovals at lower right in each panel indicate the synthesised
beam sizes (full-width at half-maximum sensitivity).  
%
}
\label{map-n2dp32}
\end{figure}

\clearpage

\clearpage

\clearpage

\clearpage

\clearpage
\begin{figure}[!t] 
\centering
\includegraphics[width=6in,angle=270]{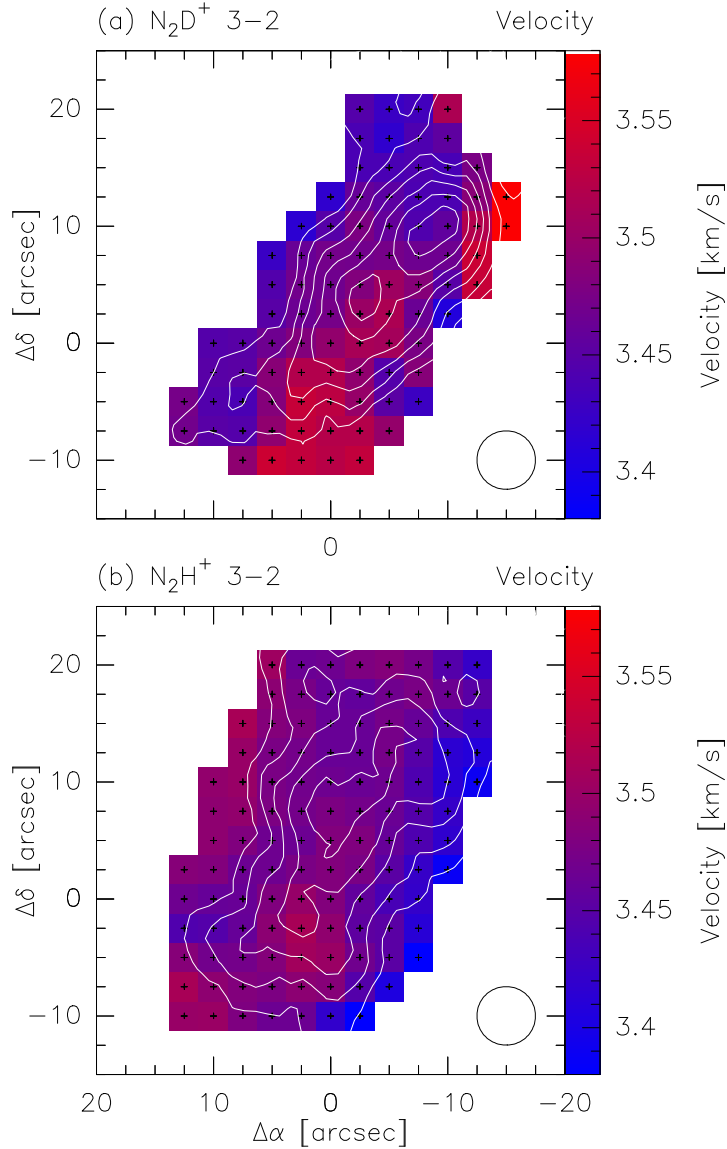}
\caption{
(a) Velocity map from hyperfine fits to the \nnd\ 3-2 data.  The data have
been resampled onto a 2\farcs5 grid with a 5\arcsec\ circular beam (shown
at lower right).  The positions where a fit was performed are indicated by
crosses.  Typical uncertainties in the fitted velocities are 0.01-0.02
\kms. The contours represent
the integrated intensity map of \nnd\ and are the same as shown in
Figure~\ref{map-n2dp32}.
(b) Velocity map from hyperfine fits to the \nnh\ 3-2 data.  Typical
uncertainties in the fitted velocities are 0.005-0.02 \kms.  Other details
are the same as in (a), 
except that the contours represent the integrated
intensity map of \nnh\ and are the same as shown in Figure~\ref{map-n2dp32}.
The origin of the maps is the position of the peak integrated \nnh\ 1-0
emission (DAM04).  The open circle indicates the beam size.
{\bf [COLOUR FIGURE]}
%
}
\label{fig-velmap}
\end{figure}

\clearpage
\begin{figure}[!t] 
\centering
\includegraphics[width=6in,angle=270]{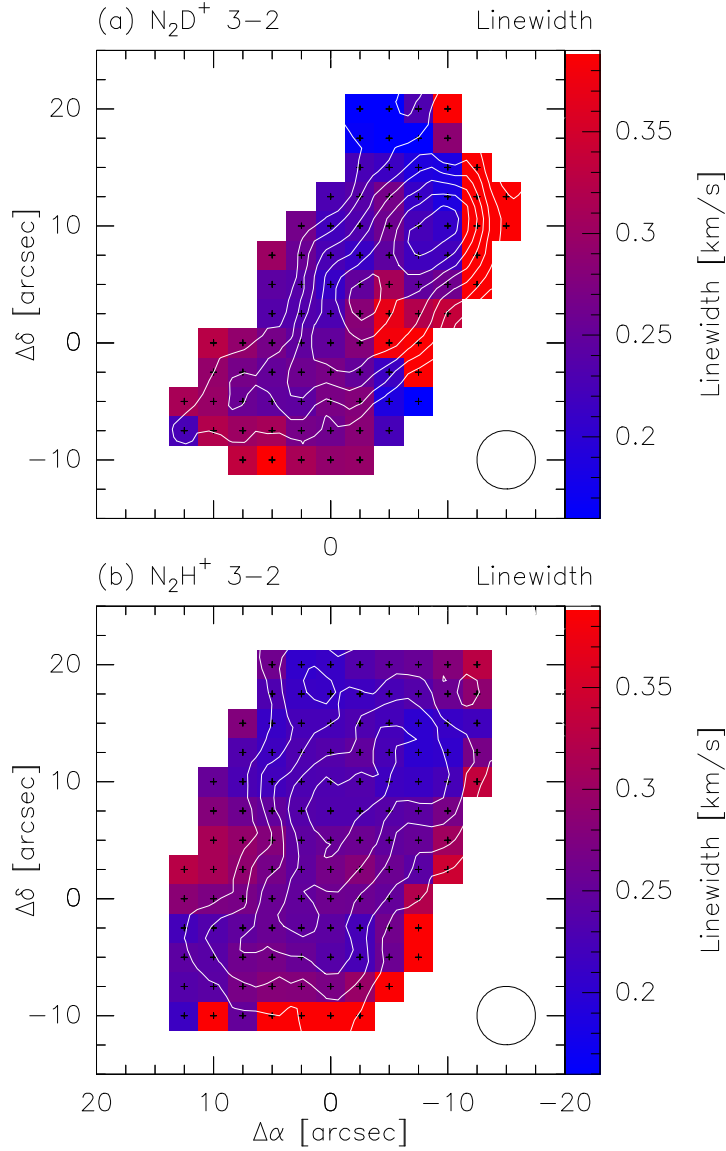}
\caption{
(a) Linewidth map from hyperfine fits to the \nnd\ 3-2 data.  The data have
been resampled onto a 2\farcs5 grid with a 5\arcsec\ circular beam (shown
at lower right).  Typical uncertainties in the fitted linewidths are
0.02-0.04 \kms.  The positions where a fit was performed are indicated by
crosses.  The contours represent
the integrated intensity map of \nnd\ and are the same as shown in
Figure~\ref{map-n2dp32}.
(b) Linewidth map from hyperfine fits to the \nnd\ 3-2 data.  Typical
uncertainties in the fitted linewidths are 0.01-0.03 \kms.  Other details
are the same as in (a), except that the contours represent the integrated
intensity map of \nnh\ and are the same as shown in Figure~\ref{map-n2dp32}.
The origin of the maps is the position of the peak integrated \nnh\ 1-0
emission (DAM04). 
{\bf [COLOUR FIGURE]}
%
}
\label{fig-linemap}
\end{figure}

\clearpage
\begin{figure}[!t]
\centering
\includegraphics[height=6in,angle=-90]{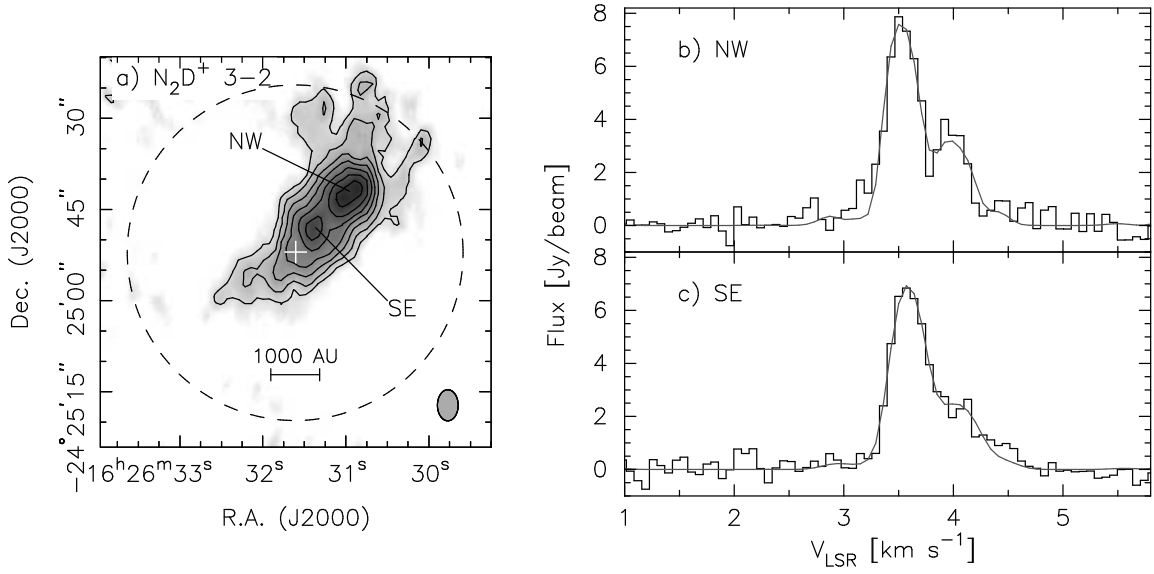}
\caption{Spectra at the integrated line map peaks of \nnd\ 3-2.  (a) The
integrated map as shown in Figure~\ref{map-n2dp32}, with the locations of
the two spectra indicated.  The large dashed circle indicates the SMA
primary beam size, while the small grey oval indicates the synthesised beam
size.  The white cross is the position of the peak integrated \nnh\ 1-0
emission (DAM04). (b) The spectrum at position NW (histogram), with a model
fit of the \nnd\ 3-2 hyperfine structure (continuous line in grey).  (c) The
spectrum at position SE (histogram), with a model fit of the \nnd\ 3-2
hyperfine structure (continuous line in grey).
%
}
\label{spec-n2dp32}
\end{figure}

\clearpage
\begin{figure}[!t] 
\centering
\includegraphics[width=6in,angle=270]{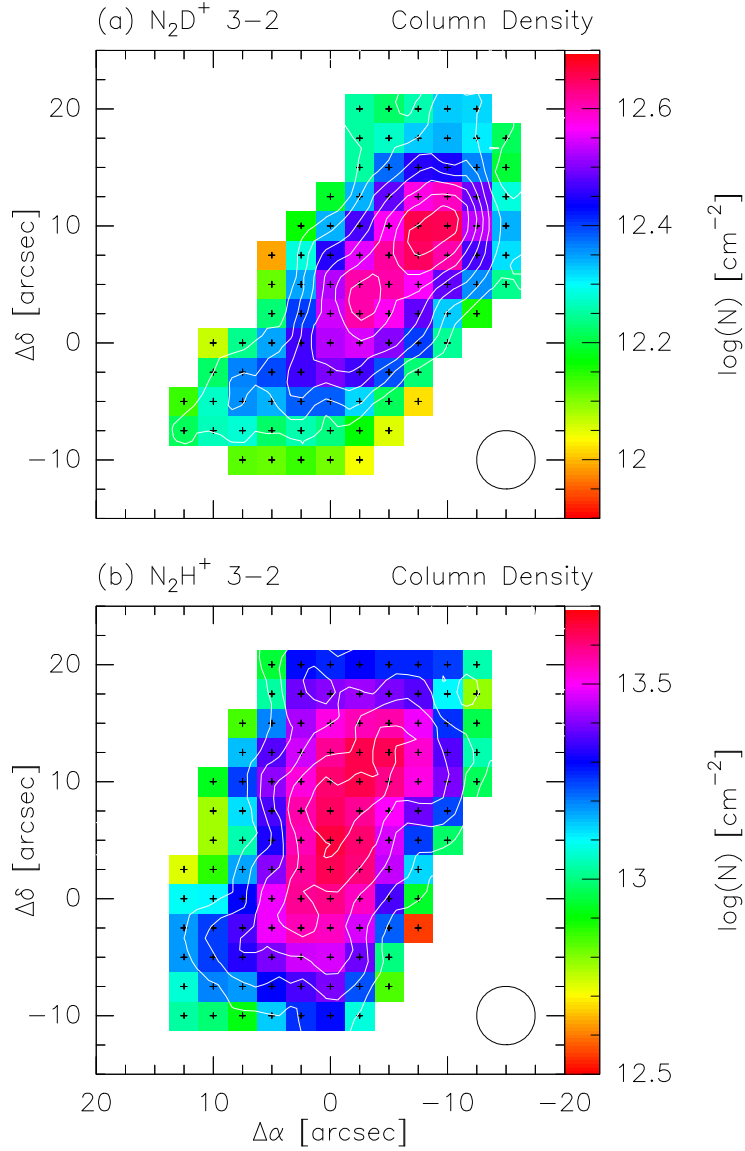}
\caption{
(a) Column density of \nnd, assuming a constant excitation temperature of
10 K.  The data have been resampled onto a 2\farcs5 grid with a 5\arcsec\
circular beam (shown at lower right).  The positions where the column
density was determined are indicated by crosses.  The contours represent
the integrated intensity map of \nnd\ and are the same as shown in
Figure~\ref{map-n2dp32}.  
(b) Column density of \nnh, assuming a constant excitation temperature of
10 K.  Other details are the same as in (a), except that the contours
represent the integrated intensity map of \nnh\ and are the same as shown in
Figure~\ref{map-n2dp32}.  The origin of the maps is the position of the
peak integrated \nnh\ 1-0 emission (DAM04).
{\bf [COLOUR FIGURE]}
}
\label{fig-coldens}
\end{figure}

\clearpage
\begin{figure}[!t] 
\centering
\includegraphics[width=6in,angle=270]{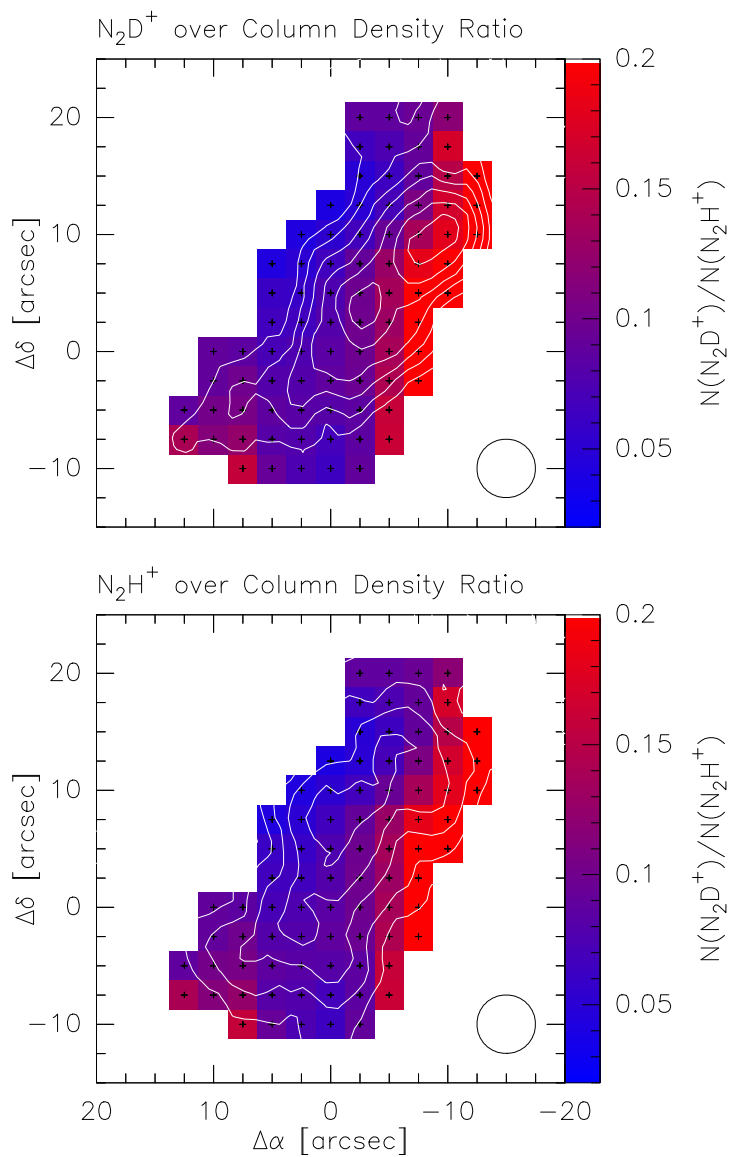}
\caption{
(a) Ratio of \nnd\ to \nnh\ column densities, using the results presented
in Figure~\ref{fig-coldens}.  The contours are integrated intensity of
\nnd\ and are the same as shown in Figure~\ref{map-n2dp32}.
(b) Ratio of \nnd\ to \nnh\ column densities, using the results presented
in Figure~\ref{fig-coldens}.  The contours are integrated intensity of
\nnh and are the same as shown in Figure~\ref{map-n2dp32}.
The resolution of the observations is
indicated by the circular beam (shown at lower right).  The origin of the
maps is the position of the peak integrated \nnh\ 1-0 emission (DAM04).
{\bf [COLOUR FIGURE]}
}
\label{fig-ratio}
\end{figure}

\clearpage

\clearpage

\end{document}